\documentclass[%
 reprint,
nofootinbib,
 amsmath,amssymb,
 aps,
]{revtex4-2}
\usepackage{soul}
\usepackage{aas_macros}
\usepackage{verbatim}
\usepackage{graphicx}
\usepackage{dcolumn}
\usepackage{bm}
\usepackage{mhchem}
\usepackage{xcolor}
\usepackage{hyperref}
\usepackage[english]{babel}
\usepackage[autostyle, english = american]{csquotes}
\MakeOuterQuote{"}

\begin{document}

\preprint{APS/123-QED}

\title{Exploring pathways to forming twin stars}

\author{Mahdi Naseri}%
 \email{mahdinaseri@arizona.edu}
\affiliation{Department of Astronomy, University of Arizona, Tucson, AZ, USA}

\author{Gabriele Bozzola}%
 \email{gbozzola@caltech.edu}
\affiliation{Department of Astronomy, University of Arizona, Tucson, AZ, USA \\
Division of Geological and Planetary Sciences, California Institute of Technology, Pasadena, CA, USA}

\author{Vasileios Paschalidis}%
 \email{vpaschal@arizona.edu }
\affiliation{Departments of Astronomy and Physics, University of Arizona, Tucson, AZ, USA}

\date{\today}

\begin{abstract}
A viable model for the dense matter equation of state above the
nuclear saturation density includes a hadron-to-quark phase transition at
densities relevant to compact objects. In this case, stable hybrid
hadron-quark stars can arise. An even more interesting scenario is one
where the hadron-to-quark phase transition results in the emergence of
a third branch of stable compact objects (in addition to white dwarfs
and neutron stars). Inherent to the presence of a third family of
compact stars is the existence of twin stars -- hybrid stars with the
same mass as the corresponding neutron stars, but with smaller
radii. Interestingly, the neutron star-twin star scenario is
consistent with GW170817. If twin stars exist in nature, it
raises a question about the mechanism that leads to their
formation. Here, we explore gravitational collapse as a pathway to the
formation of low-mass twin stars. We perform fully general
relativistic simulations of the collapse of a stellar iron core,
modeled as a cold degenerate gas, to investigate whether the end
product is a neutron star or a twin star. Our simulations show that
even with unrealistically large perturbations in the initial
conditions, the core bounces well below the hadron-to-quark phase
transition density, if the initial total rest mass is in the twin star
range. Following cooling, these configurations produce neutron
stars. We find that twin stars can potentially form due to mass loss,
e.g., through winds, from a slightly more massive hybrid star that was
initially produced in the collapse of a more massive core or if the
maximum neutron star mass is below the Chandrasekhar mass limit. The challenge
in producing twin stars in gravitational collapse, in conjunction with
the fine-tuning required because of their narrow mass range, suggests
the rarity of twin stars in nature.

\end{abstract}

\maketitle


\section{\label{sec:level1}Introduction}

One of the most important open questions in nuclear physics research
is the state of matter above the nuclear saturation density,
$\rho_s\approx2.7\times 10^{14}$ g cm$^{-3}$. The equation of state
(EOS) of cold neutron-rich matter is reasonably well understood at
densities below $\sim 2\,\rho_s$~\cite{Lattimer:2012nd,
  Drischler:2020hwi,Somasundaram:2022ztm} (see
also~\cite{Epelbaum:2008ga, Machleidt:2011zz, Tews:2012fj,
  Hebeler:2013nza, Drischler:2013iza, Hebeler:2015hla, Holt:2016pjb,
  Lonardoni:2019ypg, Drischler:2021kxf,
  Alford:2022bpp}). Uncertainties exist even for finite-temperature
dense nuclear matter; see,
e.g.,~\cite{Raithel:2021hye,Raithel:2023zml,Raithel:2023gct}.
Theoretical understanding at these densities is aided by laboratory
experiments involving heavy-ion collisions; see, e.g.,~\cite{Li:1997tz,
  Tsang:2012se, Horowitz:2019piw, Huth:2021bsp, Most:2022wgo,
  Li:2022cfd,Sorensen:2023zkk}. However, the dense-matter cold EOS
remains highly uncertain at densities of $\sim 2-40\,\rho_s$ because of limitations in theoretical or lattice
quantum chromodynamic (QCD) approaches; see,
e.g.,~\cite{Machleidt:2016rvv, Tews:2018kmu,Nagata:2021ugx,
  Lattimer:2021emm, Somasundaram:2022ztm}. As a result, a plethora of
theoretical possibilities for how matter behaves at densities higher
than $2\,\rho_s$ is currently allowed. In this work we focus on the
possibility that a quark deconfinement phase transition can take place
at densities encountered inside compact objects.

One of the fundamental consequences of QCD is color confinement. In
the absence of strong medium effects, quarks and gluons are confined
inside hadrons. Quark confinement can be violated at high temperatures
or high densities,  leading to the formation of a new state
of matter known as quark-gluon plasma~\cite{Shuryak:1980tp,
  McLerran:1986zb}. It is currently unclear whether quark
deconfinement can take place in the deep interiors of compact objects,
but this possibility has been considered by a number of works (see,
e.g.,~\cite{Collins:1974ky, Heiselberg:1999mq, Blaschke:2013ana,
  Benic:2014jia, Paschalidis:2017qmb, Ranea-Sandoval:2019miz,
  Bozzola:2019tit,Most:2018eaw, Bauswein:2018bma,
  Chatziioannou:2019yko, Otto:2019zjy, Gieg:2019yzq, Xie:2020rwg,
  Bauswein:2020aag, Prakash:2021wpz, Tan:2021ahl,
  Takatsy:2023xzf,Bauswein:2018bma, Most:2018eaw, Weih:2019xvw,
  Hanauske:2021rjk, Prakash:2021wpz, Haque:2022dsc, Blacker:2023afl}.

\begin{figure*}
\includegraphics[width=\linewidth]{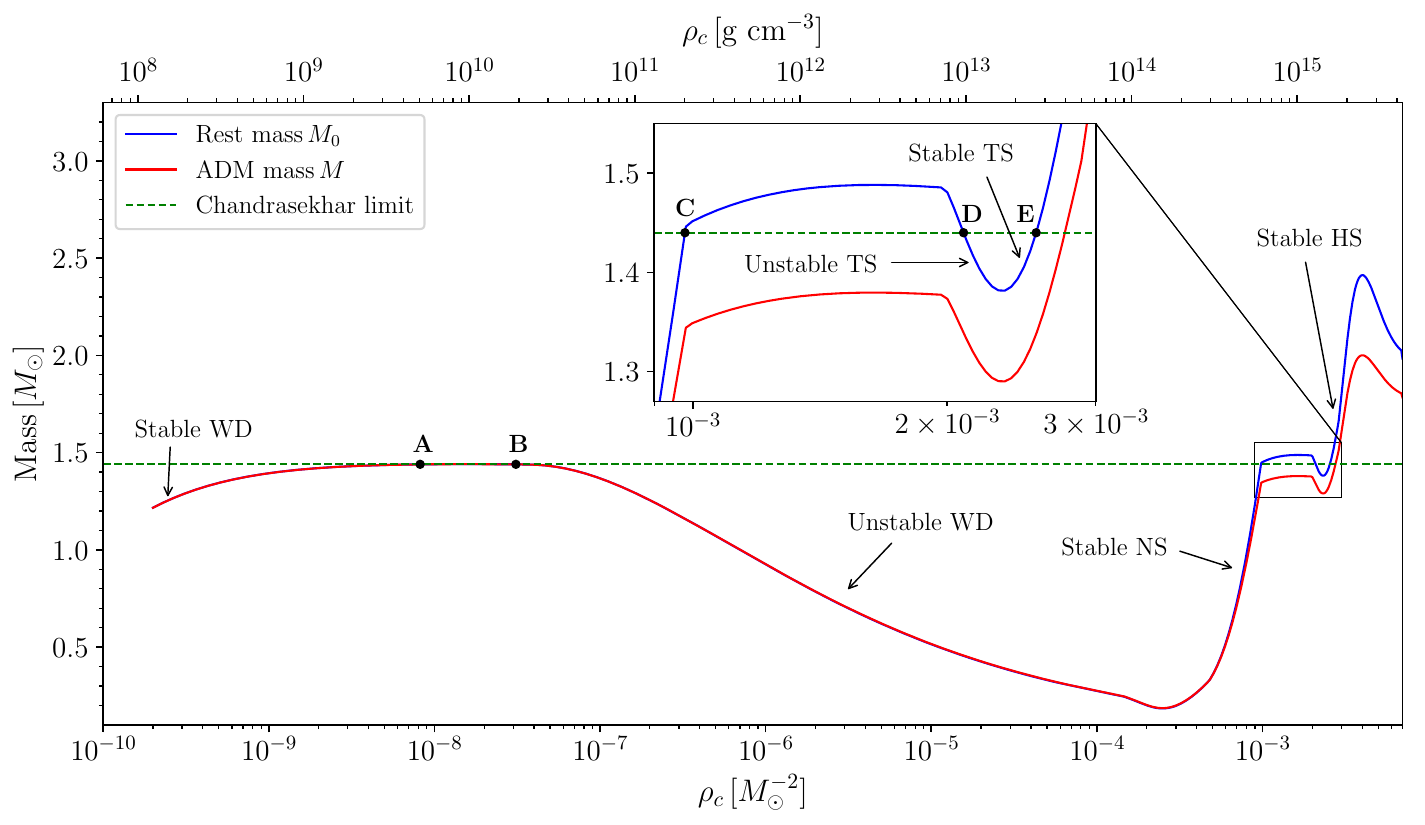}
\caption{\label{fig:wideEOS} Mass-central density plot for a sequence of TOV stars based on the EOS detailed in
  Table~\ref{EOS_table}. Both rest mass and gravitational (ADM) mass are illustrated. Stars with a fixed rest mass $M_0=M_{0,s}\sim M_{ch}$ are designated by labels \textbf{A}, \textbf{B},
  \textbf{C}, \textbf{D}, and \textbf{E}. The inset enlarges the twin star part of the curve. More details on this plot are provided in Sec.~\ref{sec:EOS}.}
\end{figure*}

If quark deconfinement takes place at astrophysically relevant
densities, an interesting scenario arises when the phase transition
occurs such that a ``third family'' of compact objects (in addition to
white dwarfs and neutron stars) emerges~\cite{PhysRevGerlach,
  Kampfer:1981yr, Glendenning:1998ag, Schertler:2000xq,
  Ayvazyan:2013cva, Zacchi:2015oma, Bejger:2016emu,
  Kaltenborn:2017hus, Alford:2017qgh, Blaschke:2018mqw,
  AlvarezCastillo:2018pve, Blaschke:2020qqj, Pfaff:2021kse}. Compact
objects in this third family are hybrid hadron-quark stars or hybrid
stars -- configurations with a quark core enveloped by a hadronic
shell. The stable third family of compact stars is separated from the
branch of stable neutron stars (NSs) by an unstable branch of compact
objects, just like stable NSs are separated from stable white dwarfs
(WDs) by an unstable branch. Fig.~\ref{fig:wideEOS} depicts these
branches on a mass versus central density plot (further details on
this plot are discussed in Sec.~\ref{sec:EOS}). As a result, there
exists a set of stable hybrid stars (HSs) that have the same mass as
NSs but smaller radii. These HSs are typically referred to as twin
stars (TSs)~\cite{Schertler:2000xq, Zacchi:2016tjw,
  AlvarezCastillo:2016oln, Christian:2017jni, Montana:2018bkb,
  Sieniawska:2018zzj, Sharifi:2021ead, Sen:2022lig,
  Tsaloukidis:2022rus}. A number of recent works
suggest that current experimental and multimessenger observations of
NSs are either consistent with or provide evidence in favor of the
existence of hadron-to-quark phase transitions in their
cores~\cite{Annala:2019puf,Annala:2021gom,Li:2022ivt,Essick:2023fso,Yamamoto:2023osc,Lin:2023cbo},
but some works do not favor this possibility~\cite{Somasundaram:2021clp,Brandes:2023hma}.  The third
family of compact objects with potentially low-mass TSs has been shown
to be consistent with a number of experiments and
observations~\cite{Paschalidis:2017qmb, Drago:2017bnf, Burgio:2018yix,
  Montana:2018bkb,
  Essick:2019ldf,DePietri:2019khb,Gorda:2022lsk,2024arXiv240102198J}.
Twin stars are particularly interesting because proof of their
existence would provide valuable insights into the quark-gluon phase
diagram.

The third family can arise when the EOS features a strong hadron-quark
phase transition (i.e.\ characterized by a sufficiently large jump in
energy density at a roughly fixed pressure) which is first order
(meaning that the first derivative of free energy with respect to a
thermodynamic variable is
discontinuous)~\cite{Glendenning:1998ag}. Alternatively, hybrid EOSs
with mixed pasta phases can also lead to the formation of the third
family; see, e.g.,~\cite{Maslov:2018ghi}. What determines the nature of the
phase transition is the surface tension between the quark and hadron
phases~\cite{glenden_book,weber_book}. If the tension between these
phases is low, a mixed phase of quark and nucleonic matter forms
in between purely nuclear and quark matter phases. By contrast, if the
tension is high, a sharp transition boundary is favorable. Both
possibilities are theoretically allowed, because the surface tension
is not known accurately.

Given the large number of works suggesting that hybrid and twin stars
are compatible with observations, an important question arises: If
these stars exist, how can they form? We begin to tackle this question
in this work.

Stellar gravitational collapse appears to be a natural pathway for the
formation of HSs. One might expect this, given that WDs and NSs are
end points of stellar evolution; see, e.g.,~\cite{Shapiro:1983du,
  Kippenhahn:2012qhp}. However, forming TSs through
gravitational collapse faces an important challenge; if one imagines
the collapse of the stellar core at approximately constant rest mass,
following cooling, the configuration would contract and encounter
first the branch of stable NSs, and then the branch of TSs,
assuming further contraction. Thus, it is unclear whether there exists
a natural pathway during stellar collapse that would prefer to form
TSs over NSs. In fact, the work of~\cite{Espino:2021adh}
treated the evolution of configurations in the TS mass range that
are initially on the unstable branch, and found that these stars
naturally transition to the stable NS branch. Thus,
Ref.~\cite{Espino:2021adh} concluded that it would be challenging to
form TSs through gravitational collapse because in a
quasiadiabatic, constant rest-mass contraction of a proto-NS that is
on its way to forming a TS, it would have to go through the
unstable branch first, which appears to favor NSs.

The primary objective of this paper is to explore whether HSs and, in
particular, TSs, can form as a result of the gravitational collapse of
a stellar core. To investigate this pathway toward HS formation, we
perform three-dimensional hydrodynamic simulations of
unstable WDs in full general relativity by considering several types
of initial perturbations differing in the degree of violence of the
collapse. The advantage of considering a WD for our initial conditions
is that it models two distinct channels to forming neutron/hybrid
stars: (i) the collapse of a degenerate iron core of a massive star
and (ii) the accretion-induced collapse of a
WD~\cite{1980ApJ...237L..81C}. Here, we do not treat detailed
microphysics or neutrino effects. Instead, we focus on treating
properly relativistic gravitation, which is critical when compact
objects arise.

Our simulations demonstrate that for a type of EOS which gives rise
to a third family of compact objects, the gravitational collapse of a
WD in the twin star  mass regime prefers the formation of NSs and not
TSs, even under extreme initial perturbations. The only pathway to
forming TSs that we were able to identify is as follows: First, a
massive core should collapse to a slightly more massive stable HS,
which may subsequently lose a small amount of mass, for example, in
the form of winds or due to rotation or a ``grazing'' collision with a
black hole, to ultimately settle as a TS. The results of our
simulations, in conjunction with the narrow mass range over which twin
stars exist suggest that twin stars should be rare. Thus, if a HS star
was involved in GW170817, then it likely was not a twin star.

The paper is structured as follows. In Sec.~\ref{sec:EOS}, we
construct the EOS that we employ in this work based on a realistic EOS
with a hadron-to-quark phase transition. In Sec.~\ref{sec:numerical},
we discuss our initial data and evolution methods along with the
numerical scheme employed in the simulations. Results of various
simulations exploring the formation of HSs under different conditions
are presented and discussed in Sec.~\ref{sec:results}. We conclude in
Sec.~\ref{sec:conclusion}, where we summarize our main
findings. Unless specified otherwise, throughout this paper we adopt
geometrized units where $c=G=1$, with $c$ being the speed of light in
vacuum and $G$ representing the gravitational constant.

\section{\label{sec:EOS}Equations of State}

There exists a broad range of possibilities for the EOS in the deep
interior of a NS. Numerous studies have explored different hadronic
models, quark models and hybrid hadron-quark models. For comprehensive
reviews readers are referred to~\cite{Lattimer:2021emm,
  Burgio:2021vgk, Kojo:2020krb, Baym:2017whm}. Apart from microscopic,
so-called realistic EOSs, the EOS can also be treated
phenomenologically, e.g., by giving the pressure or sound speed as a
function of rest-mass or energy density, see,
e.g.,~\cite{Read:2008iy,Lindblom:2010bb,Annala:2019puf,OBoyle:2020qvf}. Here,
we restrict our discussion to the astrophysical implications of EOSs
that are based on realistic models but treated phenomenologically. The
EOSs we consider give rise to a third family of stable compact
stars. However, to simulate the collapse of a stellar iron core or a
WD, the EOSs we construct must encompass a wide range of densities,
spanning the entire range from below neutron drip to the supranuclear
regime. Here we describe how we construct the phenomenological EOSs we
adopt.

The EOS below the nuclear saturation density down to WD densities is
reasonably well understood. Our base phenomenological EOS selected for
this density regime is derived from the six-parameter piecewise
polytropic EOS introduced in~\cite{Paschalidis:2010dh}. For the
high-density regime, the EOS needs to incorporate a phase transition,
so that a stable HS branch exists. The high-density EOS that we adopt
here is based on the T9 EOS of~\cite{Espino:2021adh}, which is a
piecewise polytropic representation of the T9 EOS used
in~\cite{Bozzola:2019tit}, which was, in turn, based on the ACS-II with
$\xi=0.90$ EOS of~\cite{Paschalidis:2017qmb}. This EOS describes
zero-temperature matter in beta equilibrium. Its hadronic part is
derived from~\cite{Colucci:2013pya}, the quark phase builds upon the
MIT bag model~\cite{PhysRevD.30.272, 1986A&A...160..121H,
  1986ApJ...310..261A, Zdunik:2000xx}, and the low-density crust
component is added from the models in~\cite{1973NuPhA207298N,
  1971ApJ170299B}. The quark phase of the original EOS adopts the
constant sound speed parametrization
\begin{equation} \label{T9}
P(\epsilon) = \begin{cases}
    P_{tr} & \epsilon_q\leq \epsilon \leq \epsilon_{tr} \\
    P_{tr}+c_s^2 (\epsilon - \epsilon_{tr}) & \epsilon \geq \epsilon_{tr} \; ,
\end{cases}
\end{equation}
where $\epsilon$ is the energy density,
$P_{tr}=P(\epsilon_q)=P(\epsilon_{tr})$ is the pressure of the
hadronic matter in the transition region $\epsilon \in
[\epsilon_q,\epsilon_{tr}]$, and $c_s$ denotes the sound speed. 

Given that the original T9 EOS is provided in tabulated form, we represent the high-density EOS, similar to the low-density regime, as a piecewise polytrope of the form
\begin{equation} \label{piecewise}
P = k_i \rho_0^{\Gamma_i} \; ,
\end{equation}
where $\rho_0$ is rest-mass density, and $\rho_{0,i} \leq \rho_0 \leq
\rho_{0,{i+1}}$ is the density range of each polytropic
segment. Piecewise polytropes are frequently employed in the context
of the NS EOS, as many proposed tabulated, realistic nuclear EOSs can
be well fitted by them; see, e.g.,~\cite{Read:2008iy, Hebeler:2013nza,
  AlvarezCastillo:2017qki, OBoyle:2020qvf}.

The next step is to merge these two EOSs into one that can describe
the entire range of densities relevant to compact objects. The
low-density and high-density EOSs represented as $P=P(\rho_0)$
intersect at a matching density. The combined EOS should initially
follow the low-density EOS and transition to the high-density EOS as
the density increases beyond the matching density. To ensure a smooth
transition below and above the matching density, in the density range
of approximately $10^{11}-10^{12} \textrm{g cm}^{-3}$, we linearly
interpolate between the two regimes using logarithmic pressure and
density. This step yields a base EOS that covers a wide range of
densities, extending from the crust of a WD to the dense core of a
HS\@.

After we join the low- and high-density EOSs, we solve the
Tolman-Oppenheimer-Volkoff (TOV) equations~\cite{Shapiro:1983du} for
determining general relativistic hydrostatic equilibrium stellar
configurations. We then make small modifications to the EOSs by fine-tuning the free parameters of the EOS to
satisfy the following set of conditions:

\begin{enumerate}
\item A third family of compact objects is present.
  
    \item A $2M_\odot$~\cite{Demorest:2010bx, Antoniadis:2013pzd,
      Fonseca:2016tux, NANOGrav:2019jur, Fonseca:2021wxt} compact
      object should be allowed by the EOS.

    \item The Chandrasekhar mass $M_{ch}=1.44 M_\odot$ is the maximum
      WD mass, assuming a mean molecular weight of electrons
      $\mu_e=2$.
      
    \item For the range of densities in our simulations, the sound
      speed should be subluminal. The sound speed for a
      piecewise polytropic EOS is calculated as~\cite{Read:2008iy}
    \begin{equation} \label{sound}
    c_s=\sqrt{\frac{dP}{d\epsilon}}=\sqrt{\frac{\Gamma_i P}{\epsilon + P}} \; .
    \end{equation}

    \item In the phase transition region, the sound speed should not be
      zero because the equations of hydrodynamics would be only weakly
      hyperbolic (see~\cite{Espino:2021adh} and discussion therein).
    
    \item The dominant energy condition should be satisfied for all
      densities. In the case of a perfect fluid, this
      implies~\cite{Carroll:2004st}
    \begin{equation} \label{dominant}
    \epsilon \geq |P| \; .
    \end{equation}
    
\end{enumerate} 

Following small modifications to the piecewise polytropic parameters
of our EOS such that all aforementioned conditions are met we
constructed the EOSs presented in Table~\ref{EOS_table}. The EOS
designated as ``EOS I'' in the table is used throughout most of this
study. ``EOS II'' on the other hand is used only in one our
simulations of gravitational collapse to a hybrid star above the twin
star mass range. In EOS II, the maximum mass of the NS branch is
slightly lower than $M_{ch}$, while the other characteristics are
almost identical to those of EOS I. For the remainder of this paper,
we always refer to EOS I in Table~\ref{EOS_table} unless EOS II is
explicitly specified.  A plot of these EOSs is provided in
App.~\ref{sec:EOS_app}.

The choice of piecewise polytropic parameters is not unique, but there
is not much room for changing these parameters while still meeting all
aforementioned conditions. Our goal here is not to explore all
possible equations of state, because this is not feasible. However, by
adopting the resulting phenomenological EOSs, we can perform our
point-of-principle calculations.

In geometrized units\footnote{The conversion factor between cgs units
and geometrized units with $M_\odot$ is given by
$\rho_{\text {cgs}}=G^{-3}\,M_\odot^{-2}\,c^6\,\rho_{G}$ for density, and by
$R_{\text {cgs}}=G\,M_\odot\,c^{-2}\, R_{G}$ for length, where $\rho_G$ and
$R_G$ stand for density and radius in geometrized units,
respectively.} with $M_\odot=1$, the first coefficient of the EOS in
Eq~\eqref{piecewise} is $k_1=20.7$. The value of $k_i$ for $i>1$, is
determined by continuity condition at the boundary of each two
neighboring segments, expressed as
$P_i(\rho_{0,i+1})=P_{i+1}(\rho_{0,i+1})$. This condition leads to
\begin{equation} \label{k_i}
k_{i+1}=k_i \, \rho_{0,i+1}^{\Gamma_i - \Gamma_{i+1}} \; .
\end{equation}

\begin{table}[h]
  \caption{\label{EOS_table} The two families of piecewise polytropic EOS that are adopted
    in this work, presented as eight-branch piecewise polytropes with 16
    free parameters ($\Gamma_i$, $\rho_{0,i}$, and $k_1$). Setting
    $M_\odot=1$, the first coefficient is $k_1=20.7$, and the values
    of $k_i$ for the other segments are determined by continuity at
    $\rho_{0,i}$, as given by Eq.~\eqref{k_i}. }
\begin{ruledtabular}
\begin{tabular}{l|cc|cc}
& \multicolumn{2}{c|}{\textbf{EOS I}} & \multicolumn{2}{c}{\textbf{EOS II}}\\
\toprule
\textrm{$i$}&
\textrm{$\Gamma_i$}&
\textrm{$\log_{10}\rho_{0,i}[M_{\odot}^{-2}]$}&
\textrm{$\Gamma_i$}&
\textrm{$\log_{10}\rho_{0,i}[M_{\odot}^{-2}]$}\\
\colrule
1 & 1.5000 & --- & 1.5000 & --- \\
2 & 1.3350 & -9.8833 & 1.3350 & -9.8833 \\
3 & 1.1386 & -7.4573 & 1.1286 & -7.4661 \\
4 & 2.3544 & -3.8475 & 2.3144 & -3.8475 \\
5 & 3.3458 & -3.3199 & 3.3858 & -3.3109 \\
6 & 0.2576 & -3.0081 & 2.4957 & -3.0126 \\
7 & 5.1878 & -2.7019 & 4.1878 & -2.7290 \\
8 & 7.6102 & -2.5405 & 8.1102 & -2.5405 \\
\end{tabular}
\end{ruledtabular}
\end{table} 

At sufficiently high densities, relevant for the most massive stellar
configurations, the sound speed calculated by Eq.~\eqref{sound} becomes
superluminal. Superluminality for ultradense matter with realistic
EOSs is not uncommon. For example, it arises in the
Akmal-Pandharipande-Ravenhall (APR) EOS~\cite{Akmal:1998cf}, and
efforts have been made to explain this behavior of $c_s$ in high
densities, e.g.\ in~\cite{PhysRev.170.1176}. However, we checked that
the sound speed never becomes superluminal in our dynamical spacetime
simulations.

During the process of adjusting the EOS, we find that in order to have
heavy HSs with a maximum mass of at least $2M_\odot$, the
sound speed must be large at densities corresponding to pure quark
matter. This behavior aligns closely with the findings
of~\cite{Alford:2013aca, Moustakidis:2016sab, Reed:2019ezm,
  Tews:2018kmu} on the necessity of sound speed becoming close to the
speed of light in densities of a few times $\rho_s$. Physically, this
implies that pure quark matter at intermediate densities is strongly
coupled, and violates $c_s^2=1/3$ predicted for asymptotically free
quarks when $\rho_0 \gtrsim 40 \, \rho_s$~\cite{Kurkela:2009gj}.

With the EOS available, we solve the TOV equations for a wide range of
central rest-mass densities $\rho_c$ between $10^{8}$ and $10^{16} \textrm{g
  cm}^{-3}$. We show the resulting mass-central density plot in
Fig.~\ref{fig:wideEOS}. The plot displays the total gravitational mass
$M$, also known as the Arnowitt-Deser-Misner (ADM) mass, and the total
rest mass $M_0$ of static stars as functions of their central
rest-mass density, with red and blue curves, respectively. The
branches of stable ($dM/d\rho_c>0$) and unstable ($dM/d\rho_c<0$) WDs,
NSs and HSs are explicitly identified based on the turning point
theorem~\cite{Shapiro:1983du}.

\begin{figure*}
\includegraphics[width=\linewidth]{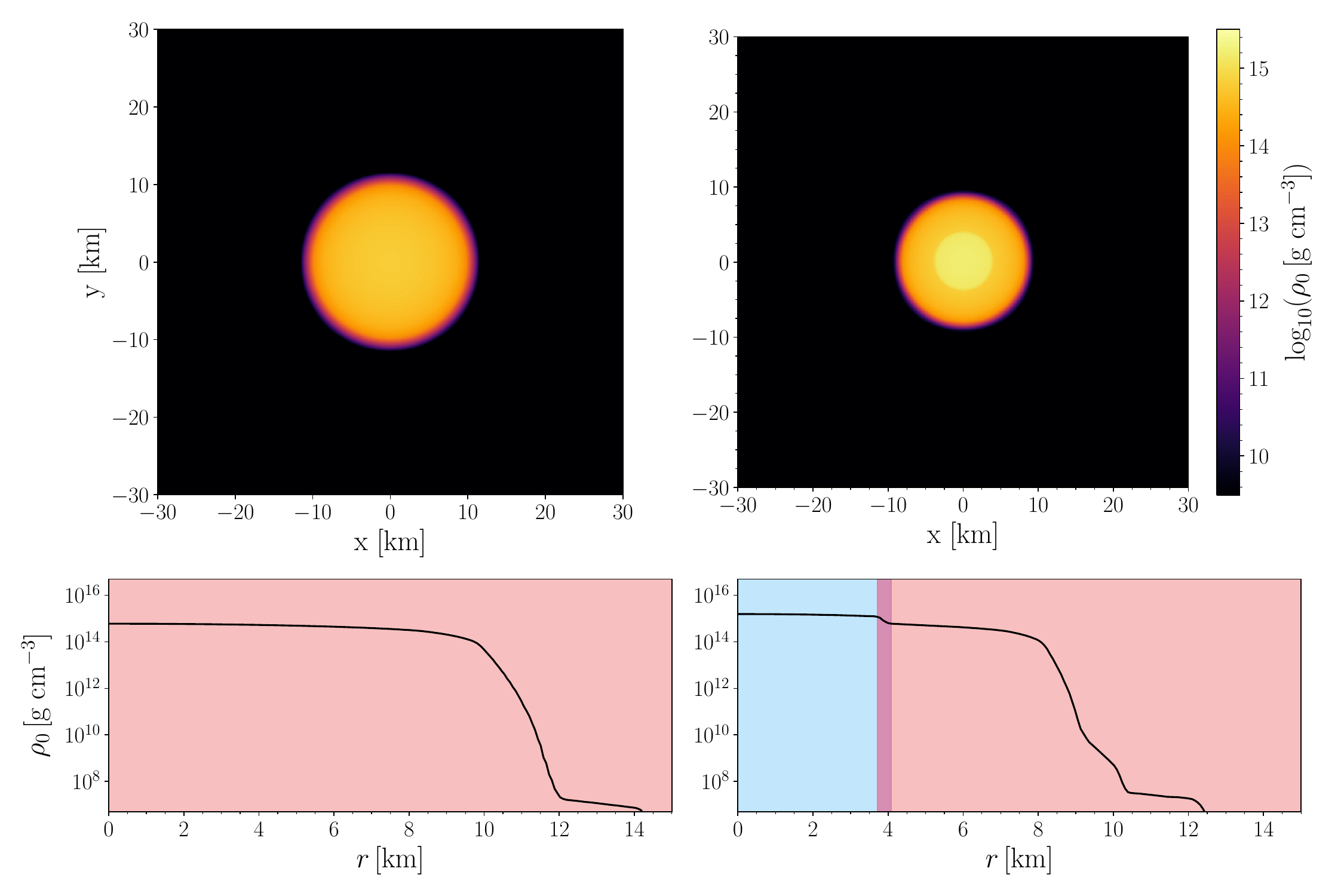}
\caption{\label{fig:Stable_NSHS} A two-dimensional view (top panels) of
  the matter distribution inside the stable NS configuration
  \textbf{C} (left) and the HS configuration \textbf{E} (right). For
  the HS, the distinct quark core is easily distinguishable by its
  brighter color, indicative of the sudden increase in density. The
  density profile of each star is also displayed in the bottom panels
  as a function of the radial coordinate $r$. Red color in the
  background stands for the pure hadronic matter, blue represents the
  pure quark matter, and the narrow purple region in the plot of the
  HS density profile denotes the zone containing a mixture of quarks
  and hadrons.}
\end{figure*}

Our initial data for the collapse simulations are based on stellar
configuration \textbf{B} in Fig.~\ref{fig:wideEOS}, with a rest mass
$M_{0,s}$ very close to $M_{ch}$ but on the unstable WD branch to help
accelerate the collapse. The mass of this star, $1.44\,M_\odot$, falls
right in the mass range of TSs and allows us to perform our
point-of-principle calculations to test for twin star formation. The
TOV solutions with this choice of mass are labeled on this plot as
$\{$\textbf{A}, \textbf{B}, \textbf{C}, \textbf{D},
\textbf{E}$\}$. Configurations \textbf{A} and \textbf{B} are a stable
and an unstable WD, respectively. Configurations \textbf{C} and
\textbf{E} represent a stable NS and a stable HS in the twin star mass
regime, respectively, although, strictly speaking, TSs have the same gravitational
mass and not the same rest mass. The internal structure of these two
compact stars is illustrated in Fig.~\ref{fig:Stable_NSHS}.

The largest ADM mass is attained in our EOSs is by a HS. Stars on the
unstable branch that separates stable NSs and stable HSs, such as
configuration \textbf{D} (see the inset in Fig.~\ref{fig:wideEOS}), which
we will refer to as unstable TSs, are entirely in the twin star mass
range. Fig.~\ref{fig:wideEOS} can help visualize the basic question in
our work. During the gravitational collapse of a stellar core/WD along
a constant rest-mass path (dashed green horizontal line in
Fig.~\ref{fig:wideEOS}), as the remnant cools it encounters first the
stable NS branch and then the stable TS branch as it contracts
quasiadiabatically. This begs the following question: If the NS branch
is encountered first, then how can stable twin stars form?

We point out that during the evolution the remnant entropy changes due
to heating; therefore, the actual evolutionary track that we describe
above is not on the plane of Fig.~\ref{fig:wideEOS}, because that
corresponds to zero-entropy configurations. However, core bounce
occurs above the density of the NS configuration~\textbf{B}, and as the remnant cools and contracts the evolutionary path approaches the
one shown by the horizontal dashed line in Fig.~\ref{fig:wideEOS}.

\begin{figure}[h]
\includegraphics[width=\linewidth]{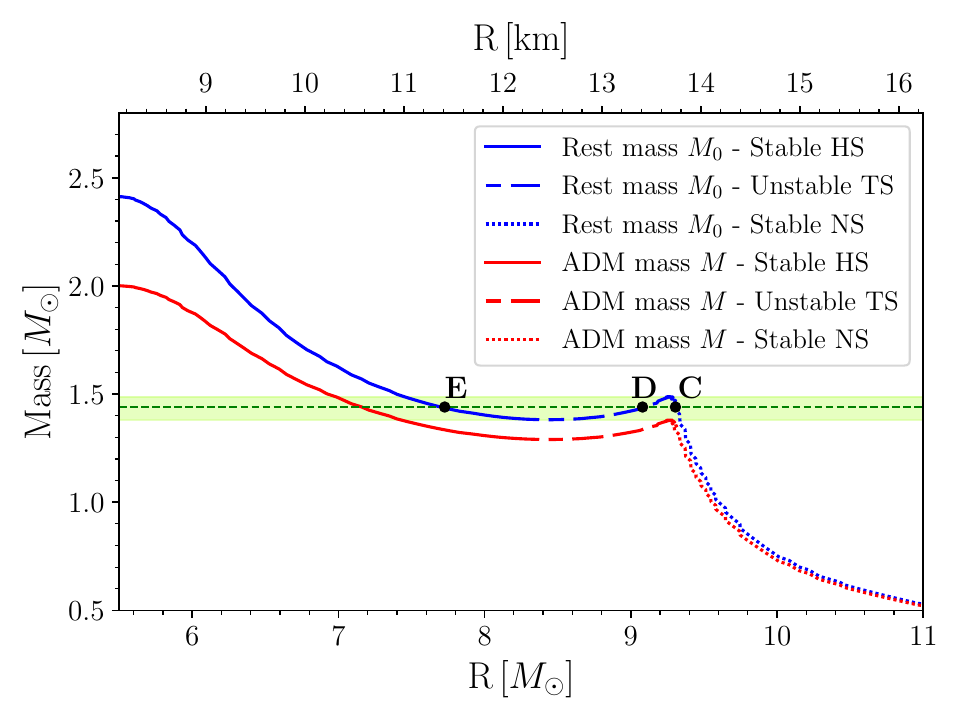}
\caption{\label{fig:MR} Mass-radius relationship based on the EOS used
  in this work. Colors and points indicate the same quantities and
  configurations as in Fig.~\ref{fig:wideEOS}. Solid, dashed, and
  dotted curves stand for stable HS, unstable TS, and stable NS
  branches, respectively. The green shaded area represents the
  mass range of TSs.}
\end{figure}

In addition to mass, the radius is another important macroscopic
property of compact objects. For any given EOS, solving the TOV
equations also determines the radius of each
configuration. Consequently, every EOS has a unique mass-radius (M-R)
relation that can be probed through multimessenger observations and,
subsequently, impose constraints on the EOS of dense nuclear matter
(see, e.g.,~\cite{Ozel:2016oaf,
  Haensel:2016pjp,Horowitz:2019piw,Raithel:2019uzi,Radice:2020ddv,Lattimer:2021emm}
for reviews). The M-R relation based on our EOS is displayed in
Fig.~\ref{fig:MR}. The three compact stars labeled in
Fig.~\ref{fig:wideEOS} are also shown in this plot with the same
labels. We show these stars on the rest-mass versus radius curve, because
$M_0$ is conserved during collapse.

\section{\label{sec:numerical} Initial data and evolution Methods}
In the $M-\rho_c$ plane of Fig.~\ref{fig:wideEOS}, the unstable WD 
configuration \textbf{B} will collapse at constant rest mass, and following
cooling, it must ultimately migrate toward one of the two stable
configurations with the same rest mass, i.e., at points \textbf{C} or
\textbf{E}. During its evolution, the star moves through configurations
that cannot be accurately described by cold equilibrium models, so \emph{a priori},
it is unclear which of these stable states the star will end up to.
To determine the ultimate fate of
the unstable WD, we perform multiple simulations of gravitational
collapse by perturbing configuration \textbf{B}. As the initial
configuration contracts, it heats adiabatically, and via shocks, if
the core bounces. Therefore, to end up on a cold degenerate
configuration, the star has to cool. We adopt a covariant local
effective cooling approach, the details of which we present in
App.~\ref{sec:cooling}. Our cooling model is that
of~\cite{Paschalidis:2011ez}, and is characterized by a single
parameter--the cooling timescale $\tau_c$.

Except for the code that builds our tabulated EOS in Python, and our
TOV solver, our computational framework is based on the \texttt{Einstein
  Toolkit}~\cite{Loffler:2011ay, EinsteinToolkit:2023_05} --
\texttt{Cactus}~\cite{Goodale2002a} and
\texttt{Simfactory}~\cite{sim}, employing \texttt{Carpet} for mesh
refinement~\cite{Schnetter:2003rb}. Postprocessing of the simulation
data is performed using the \texttt{kuibit} Python
package~\cite{Bozzola:2021hus}.

\subsection{Initial data}
The initial data are provided by the \texttt{Einstein Toolkit} thorn
\texttt{RNSID} which is based on the \texttt{RNS}
code~\cite{1995ApJ444306S, Stergioulas:2003yp}. This code builds
(rotating) isolated stellar configurations assuming a zero-temperature
EOS in either tabulated or polytropic forms. As our EOS is represented
by a piecewise polytrope~\ref{EOS_table}, we convert it to a highly
sampled table before passing it to \texttt{RNSID}. For simplicity, we
ignore rotation in this work.

To vary the degree of violence of the collapse of our initial
configuration we consider a set of different initial perturbations
that involve nonzero velocity and/or pressure perturbations. We also
consider mass depletion perturbations. We describe these below. After
we impose the initial perturbations we do not resolve the Einstein
constraints, but we check in all our cases that the constraint
violations are always small during the evolution.

\subsubsection{Pressure perturbation}
The pressure perturbations are modeled as
\begin{equation} \label{pressure_pert}
P\rightarrow (1+\xi_p)P =P+\delta P\; ,
\end{equation}
for every point in the configuration. Here, the value of $\xi_p
={\delta P}\slash{P}$ controls the strength of the perturbation. The
pressure perturbation could be either negative or positive. A negative
$\delta P$ corresponds to pressure depletion and consequently, speeds
up the collapse. On the contrary, a positive pressure perturbation
corresponds to heating the stellar interior that must be radiated away
to maintain the equilibrium. We use the latter to test our
implementation of effective cooling in App.~\ref{sec:A}.

\subsubsection{Velocity perturbation}
Our model for velocity perturbation is given by the
following two-parameter function
\begin{equation} \label{velocity_pert}
\vec{v}\rightarrow  \xi_v \, {\left(\frac{r}{R}\right)}^\kappa \; \hat{r} \; ,
\end{equation}
where $\vec{v}$ is the coordinate 3-velocity, $\hat{r}$ is the radial unit vector,
$r$ is the coordinate radius, $R$ is the coordinate radius of the star,
and $\xi_v$ determines the amplitude of the velocity perturbation and
could be either negative (collapsing star) or positive (expanding
star). The exponent $\kappa$ controls the radial profile of the
perturbation. The case $\kappa=0$ is a particular model with a
velocity perturbation independent of position. The choice of a
power-law perturbation model is made for simplicity and is not
unique. We choose $\kappa\neq0$ so the perturbation goes smoothly to
zero close to the center of the star. This is because for
$\xi_v=\mathcal{O}(-0.1)$ and $\kappa=0$ the star bounces almost
immediately and explodes.

\subsubsection{Mass perturbation}
To simulate a configuration that experiences a small amount of mass
loss, e.g., lost due to winds, the initial data are perturbed such
that a low-density shell of the stellar structure is set to the
tenuous atmospheric density we maintain in our simulations. This
perturbation essentially changes the total rest mass as
\begin{equation} \label{mass_pert}
M_0\rightarrow (1+\xi_m)M_0 =M_0+\delta M_0\; .
\end{equation}
Here, $\xi_m={\delta M_0}\slash{M_0}$ specifies the fraction of rest
mass removed from or added to the original star and, hence, takes a
negative value in the case of mass loss. In practice, this is done by
defining a threshold rest-mass density below which the rest-mass
density of the stars takes the value of the tenuous atmospheric
density until the desired amount of mass $\delta M_0$ is depleted.

The parameters characterizing all simulations
performed in this work are summarized in Table~\ref{simulations}.

\begin{table*}
\caption{\label{simulations}The set of simulations conducted in this
  study using EOS I is detailed below. For each case, the
  corresponding configuration in the $M-\rho_{c}$ plot for the initial
  data is listed, along with the perturbation parameters. The last
  column describes how cooling is performed during the simulation, if
  enabled. Further details are provided in
  Sec.~\ref{sec:results}. Configuration \textbf{F} represents a stable
  HS with $M_0=1.49 M_\odot$, and is discussed in
  Sec.~\ref{mass_loss}. The simulations based on the last two rows are
  presented in App.~\ref{sec:stable_stars}.  }
\begin{ruledtabular}
\begin{tabular}{l|cccccc}
 Description & Initial TOV solution & $\xi_p$ & $\xi_m$
& $\xi_v$ & $\kappa$ & Cooling\\ \toprule
 Unstable WD collapse & \textbf{B} & -0.01 & 0.00 & 0.00 & 0.0 & SI \\
 Unstable WD collapse & \textbf{B} & -0.01 & 0.00 & 0.00 & 0.0 & SII \\
 Unstable WD collapse --- perturbed pressure & \textbf{B} & -0.90 & 0.00 & 0.00 & 0.0 & SI \\
 Unstable WD collapse --- perturbed pressure & \textbf{B} & -0.90 & 0.00 & 0.00 & 0.0 & SII \\
 Unstable WD collapse --- perturbed velocity & \textbf{B} & -0.01 & 0.00 & -0.10 & 0.0 & --- \\
 Unstable WD collapse --- perturbed velocity & \textbf{B} & -0.01 & 0.00 & -0.10 & 0.5 & SII \\
 Unstable WD collapse --- perturbed velocity & \textbf{B} & -0.01 & 0.00 & -0.10 & 1.0 & SII \\
 Unstable WD collapse --- perturbed velocity & \textbf{B} & -0.01 & 0.00 & -0.10 & 5.0 & SII \\
 Unstable WD collapse --- highly perturbed & \textbf{B} & -0.90 & 0.00 & -0.10 & 0.5 & SII \\
 Unstable WD collapse --- highly perturbed & \textbf{B} & -0.90 & 0.00 & -0.10 & 1.0 & SII \\
 Unstable WD collapse --- highly perturbed & \textbf{B} & -0.90 & 0.00 & -0.10 & 5.0 & SII \\
 Unstable WD collapse --- highly perturbed & \textbf{B} & -0.90 & 0.00 & -0.30 & 1.0 & SII \\
 Unstable WD collapse --- highly perturbed & \textbf{B} & -0.90 & 0.00 & -0.40 & 1.0 & SII \\
 Stable massive HS & \textbf{F} & 0.00 & 0.00 & 0.00 & 0.0 & --- \\
 Massive HS --- mass loss & \textbf{F} & 0.00 & -0.03 & 0.00 & 0.0 & --- \\
 Stable NS & \textbf{C} & 0.00 & 0.00 & 0.00 & 0.0 & --- \\
 Stable HS & \textbf{E} & 0.00 & 0.00 & 0.00 & 0.0 & --- \\
\end{tabular}
\end{ruledtabular}
\end{table*}

\subsection{Evolution}
The spacetime initial data are evolved in the
Baumgarte-Shapiro-Shibata-Nakamura (BSSN)
formulation~\cite{Baumgarte:1998te, Shibata:1995we} as implemented in
the public \texttt{Lean} code~\cite{Sperhake:2006cy}. The spacetime
gauge choice adopted here includes the "1+log" and the "$\Gamma$ --
driver" conditions~\cite{Bona:1994dr, Alcubierre:2002kk}.  The
hydrodynamic initial data are evolved with the publicly available code
\texttt{IllinoisGRMHD}~\cite{Etienne:2015cea, github_GRMHD}. We have
implemented in \texttt{IllinoisGRMHD} the cooling
of~\cite{Paschalidis:2011ez} that we describe in
App.~\ref{sec:cooling} . We also test our implementation of cooling in
App.~\ref{sec:A}. The EOS for the evolution is hybrid with a cold and
thermal component
\begin{equation}\label{cold_th}
P=P_{\rm cold} + P_{\rm th},
\end{equation}
where the cold pressure $P_{\rm cold}$ is given by EOS I or II, and
$P_{\rm th}=(\Gamma_{\rm th}-1)\epsilon_{\rm th}$, with $\epsilon_{\rm
  th}$ being the thermal energy density. We employ $\Gamma_{\rm
  th}=2$.

In the WD collapse simulations, since the radius of the initial
configuration (WD) is about 100 times larger than that of the final
configuration (HS/NS), it is computationally inefficient to perform
the simulation with the highest resolution necessary to resolve a
HS/NS from the start.  Thus, we employ adaptive mesh refinement. The
initial grid structure consists of two refinement levels, with the finer
resolution set to 100 points across the WD. However, as the radius
decreases over time (and the central density increases), the initial
resolution becomes inadequate, and we add progressively higher
resolution refinement levels. Based on an estimate of the final
central density as $\rho_{c,f}\sim10^{-3} \, M_\odot^{-2}$ using the
equilibrium cold configurations (see Fig.~\ref{fig:wideEOS}) we
construct six refinement levels that are initially inactive. As the
central rest-mass density rises, these refinement levels are
subsequently activated when the density reaches certain values. When
all the refinement levels are activated, the highest grid resolution
is $166\,\textrm{m}$. The density range between the initial central
density, $\rho_{c,i}$, and $\rho_{c,f}$ is divided into six equidistant
logarithmic segments. Consequently, there are seven values of density,
separating these six segments, that we denote $\{ \rho_1, \rho_2,
\ldots, \rho_7 \}$, where $\rho_1=\rho_{c,i}$ and
$\rho_7=\rho_{c,f}$. Every time the maximum density of the star
exceeds $\rho_{k} = {\left(\frac{\rho_{c,f}}{\rho_{c,i}}\right)}^{1/6}
\times \rho_{k-1}$ for $k=\{ 2, 3, \ldots, 7 \}$, a new refinement
level is activated. In the collapse simulations, the outer boundary is
set to $1.7 R_{\rm WD}\approx 498M_{\rm WD}$ ($R_{\rm WD}=1058.55\,
\textrm{km}$ is the initial WD radius and $M_{\rm WD}$ its mass.). The
half-side length of each new refinement level is half that of the
preceding level and its resolution twice that of the preceding
level. For the other simulations in this work (e.g.\ starting with a
more massive HS), the grid has the same structure as described for the
stable HS in App.~\ref{sec:stable_stars}.

As the star collapses, it heats adiabatically, eventually resulting in
a core bounce once the EOS has stiffened and significant heat has been
generated. Without taking cooling into account, the collapse would
stall and the star would not continue to contract to reach either of
the equilibrium configurations on the TOV sequence. This excess heat
is naturally radiated away in the form of neutrinos.

In the absence of a physically accurate 6+1 dimensional neutrino code,
we crudely model cooling by locally removing any excess heat, as
described in App.~\ref{sec:cooling}. The only free parameter that
needs to be specified in this model is the cooling timescale
$\tau_c$. The value of $\tau_c$ has to be chosen such that we respect
the hierarchy of timescales in our problem, while also considering the
duration over which the simulations can be completed. If $\tau_c$ is
much larger than the dynamical timescale $t_{dyn}$, the computations
will take a very long time to complete. However, $\tau_c$ is much
smaller than $t_{dyn}$, this would result in rapid cooling, causing
significant perturbations instead of a smoother transition to the
zero-temperature remnant.  As a star collapses, the density
increases. Thus, the dynamical timescale of the star, defined as
$t_{dyn}\equiv \frac{1}{\sqrt{\rho_c(t)}}$, is not constant over time.

We follow two different strategies to activate cooling in our
simulations in order to ensure that our {\rm final} results are
invariant with the cooling strategy. In the first strategy (SI), the
collapse begins without cooling, and subsequently comes to a halt due
to core bounce and heat generation. When the star settles in this
state, we activate cooling with $\tau_c = 3\, t_{dyn}$, where
$t_{dyn}$ is determined based on the central density of the settled
configuration. In the second strategy (SII), cooling is activated from
the beginning with $\tau_c = 3\, t_{dyn}$. However, as the collapse
continues, $\tau_c$ is updated to track the changing dynamical
timescale as the maximum rest-mass density increases. We update
$\tau_c$ every time a new refinement level is added to the
configuration with the same relation $\tau_c = 3\, t_{dyn}$, where
$t_{dyn}$ is determined based on the values of $\{ \rho_1, \rho_2,
\ldots, \rho_7 \}$ discussed above.

\section{\label{sec:results}Results}

In this section, we present the results of our simulations. We begin
with the simulations of WD collapse, both without and with strong initial
perturbations. Subsequently, we present the outcomes of the simulation
involving a HS with mass loss. Furthermore, we perform a simulation to
explore the formation of a stable HS with a mass exceeding the mass
range of TSs, arising from the collapse of a WD. We present how well
the constraints are satisfied during the evolution in
App.~\ref{sec:B}.  We also perform two simulations involving the
stable NS and HS shown in Fig.~\ref{fig:Stable_NSHS} to demonstrate
the code can reliably evolve stable stars. The results of these
simulations are discussed in App.~\ref{sec:stable_stars}.

\subsection{White dwarf collapse}
A potential pathway to forming TSs is the gravitational
collapse of the iron core of a massive star or the accretion-induced
collapse of a WD. An unstable WD near the Chandrasekhar limit is an
acceptable model for either scenario. The initial central density and
radius of the unstable WD at configuration \textbf{B} are $\rho_c=2.14\times 10^{10} \textrm{g} \,
\textrm{cm}^{-3}$ and $R_{\rm WD}=1058.55\, \textrm{km}$. Following cooling,
this configuration is expected to collapse and settle into
configuration \textbf{C} or \textbf{E}, since these are the only
stable configurations that preserve the total rest mass. To speed up the
collapse, we initially deplete $1\%$ of the pressure at every point
inside the WD, that is, $\xi_p=-0.01$.

\begin{figure*}
\includegraphics[width=\linewidth]{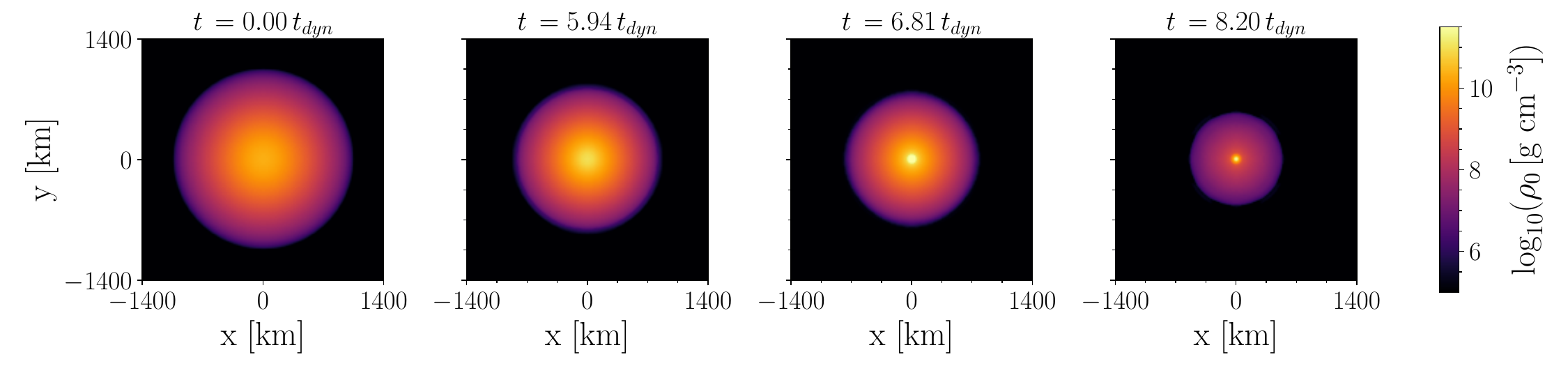}
\caption{\label{collapse_snapshots} Snapshots of matter distribution
  within the star at four different times during the collapse in the
  $\textrm{xy}$ plane. The leftmost panel shows the unstable WD configuration
  \textbf{B}, and the rightmost panel shows the stable NS
  configuration \textbf{C}. The matter left as a low-density atmosphere
  around the stable NS gradually falls onto the dense core during the
  final stages of the collapse but its total mass is negligible. Thus,
  it cannot raise the maximum density, as evidenced in
  Fig.~\ref{WDcollapse}.}
\end{figure*}

\begin{figure}[t]
\includegraphics[width=\linewidth]{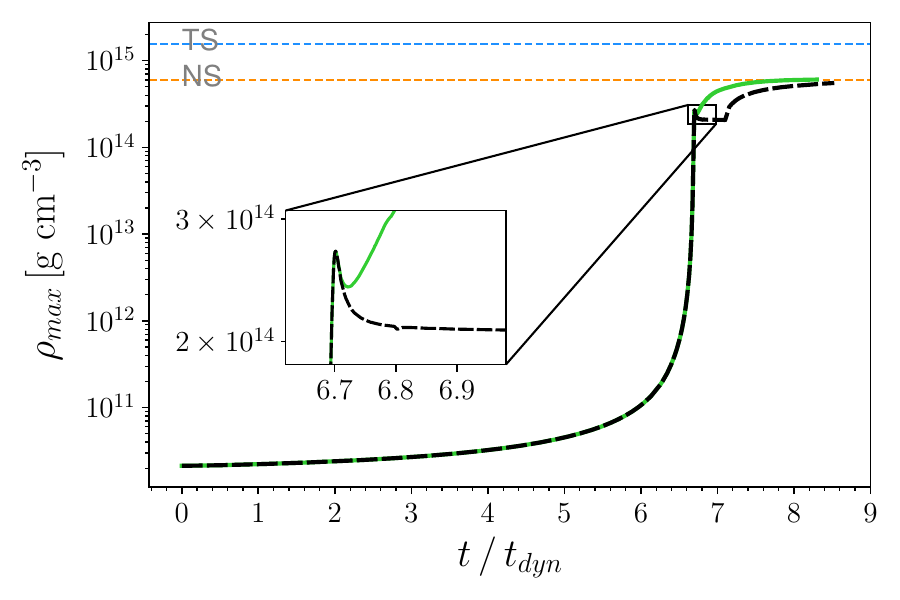}
\caption{\label{WDcollapse} Evolution of the maximum density for an
  unstable WD under 1\% pressure depletion. Here, $t_{\rm
    dyn}=26.46\,\text{ms}$ is the dynamical timescale of the initial WD. The
  orange and blue dashed lines denote the rest-mass density of the
  stable compact configurations \textbf{C} and \textbf{E} of
  Fig.~\ref{fig:wideEOS}, respectively. The black dashed (green) curve
  illustrates the evolutionary path in the SI (SII) cooling strategy.
  The inset enlarges the bounce.}
\end{figure}

Fig.~\ref{collapse_snapshots} shows the outcome of the simulation
adopting the SII cooling strategy with four snapshots of the matter
distribution. The dense core depicted in the final snapshot of this
figure shows the ultimate stable compact star. The mass in the
low-density region surrounding the core is negligible compared to that
of the central compact object. Fig.~\ref{WDcollapse} presents the time
evolution of the maximum rest-mass density inside the star. The
horizontal orange and blue dashed lines display the central rest-mass
density of the stable NS and the stable HS with the same rest mass as
the initial WD, i.e.\ configurations \textbf{C} and \textbf{E},
respectively. The black dashed (green) curve shows the evolution of
the maximum rest-mass density in the SI (SII) cooling strategy.

In both scenarios, the initial configuration slowly contracts. After
$t\sim 6\,t_{dyn}$ (here and in Fig.~\ref{WDcollapse} $t_{dyn}$
corresponds to the dynamical timescale of the initial WD), the
collapse accelerates. Around $t\sim 6.70\,t_{dyn}$, a bounce in the
density evolution occurs, halting the rapid collapse. We note that the
evolution until core bounce is insensitive to whether cooling is
active or not. After the bounce, the evolution of the maximum
rest-mass density depends on the cooling strategy. In the case where
cooling is initially inactive, the star settles at a density below
both stable NS and stable HS\@. Note that this configuration does not
lie on the TOV sequence with the underlying cold EOS, because it no
longer has zero entropy. As soon as cooling is activated, this
intermediate configuration undergoes additional collapse, and
asymptotes to the stable NS configuration \textbf{C} (orange dashed
line). In the SII case, the star continues its collapse after the
bounce, since cooling already had reduced the thermal
pressure. However, the asymptotic configuration, when most of the heat
has been removed, is again the stable NS configuration
\textbf{C}. Thus, the outcome of the collapse is independent of the
cooling strategy adopted here and the final cold configuration has a
maximum density that is below the quark-hadron phase transition
region. Therefore, the final configuration is a NS.

We note that in the part of the plot in Fig.~\ref{WDcollapse} where the green and black curves
plateau at the stable NS threshold, every unit
of $t/t_{dyn}$ corresponds to $\sim 170$ dynamical timescales of that
stable NS\@. Hence, the simulations were long enough to ensure that
the evolution has reached the stable cold configuration.

We also conducted a simulation with $\tau_c=1.5\,t_{dyn}$ as opposed
to $\tau_c=3\,t_{dyn}$, to check if the cooling timescale can affect
the results. We confirmed that faster cooling only accelerates the
collapse, while leaving the final outcome unaffected. Thus, we
conclude that the final product of the collapse of an unstable WD is a
stable NS with the same total rest mass; these initial data do not
result in a stable HS, regardless of the cooling approaches tested.

\begin{figure}[t]
\includegraphics[width=\linewidth]{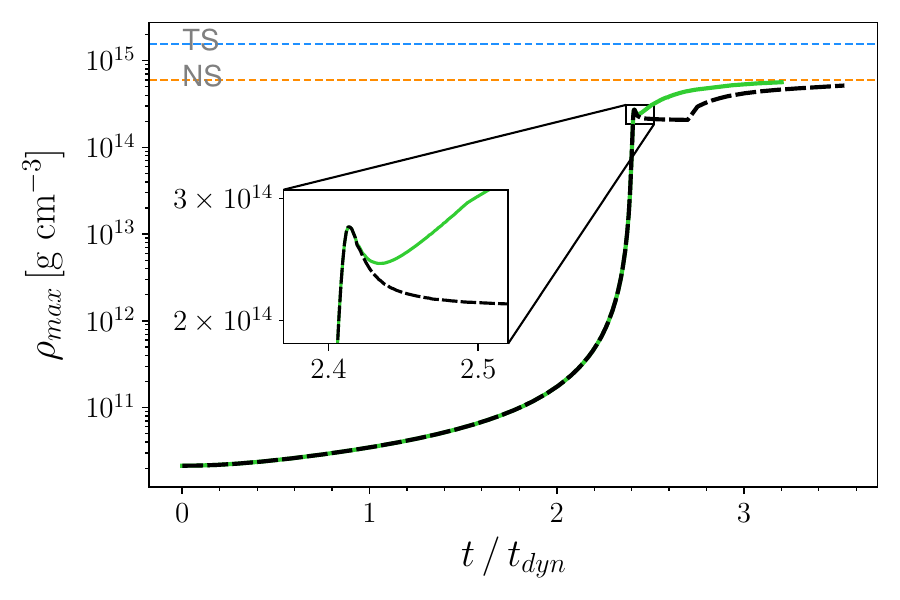}
\caption{\label{pressure_depletion} The same as Fig.~\ref{WDcollapse}, but
  for the case with 90 percent of pressure depletion in the initial
  data. The line styles and colors are coded as in
  Fig.~\ref{WDcollapse}, and similarly, $t_{\rm dyn}$ is the dynamical
  timescale of the initial WD}.
\end{figure}

\subsection{Strongly perturbed initial data}
Here we explore more violent perturbations to investigate whether the
final remnant can be a TS if the strong bounce can be avoided or
happen at a higher density past the hadron-to-quark phase
transition. To cause the bounce to occur at higher densities, one can
potentially think of more extreme initial conditions for the unstable
WD\@.

\begin{figure*}
\includegraphics[width=\linewidth]{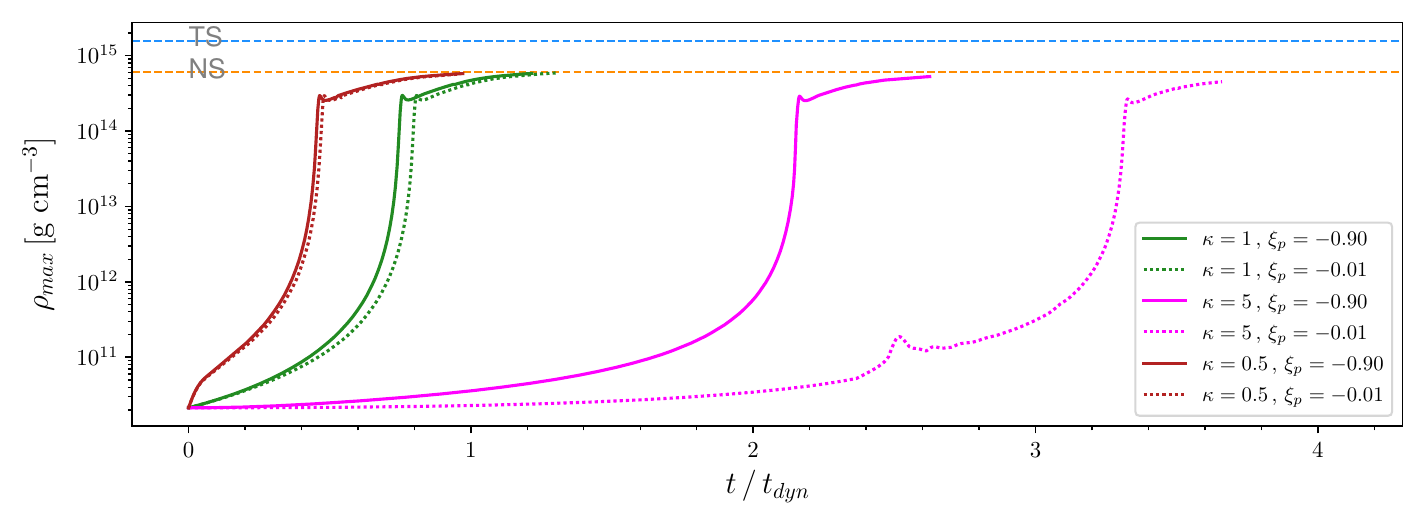}
\caption{\label{kappa} Maximum density versus time for the simulations of
  WD implosion with initial perturbations in both velocity and
  pressure. Time is scaled with the dynamical timescale of the initial
  WD. The colors maroon, green, and magenta designate the three values
  of $\kappa = \{ 0.5,1,5 \}$, respectively. The velocity perturbation
  is fixed at $\xi_v=-0.1$. Dotted lines correspond to $\xi_p=-0.01$,
  while solid lines to $\xi_p=-0.90$. All six simulations employ the
  SII cooling strategy. The dashed lines designate the central density
  of the stable NS/HS as in Fig.~\ref{WDcollapse}.}
\end{figure*}

\subsubsection{Large pressure depletion}
One of the parameters that could potentially facilitate the collapse
is the strength of the pressure depletion characterized by $\xi_p$. A
very large pressure depletion might result in a more rapid
collapse. We conduct a set of simulations with SI and SII cooling
beginning with configuration \textbf{B} but $\xi_p=-0.90$. While this
level of pressure depletion is unlikely in nature, we use it as an
extreme case to determine if collapse alone can form a TS. Since the
WD is basically a Newtonian object, such a large pressure depletion
has a negligible impact both on the total energy budget of the star
and on the constraints.

The results of these simulations are displayed in
Fig.~\ref{pressure_depletion}. By comparing
Fig.~\ref{pressure_depletion} and Fig.~\ref{WDcollapse}, it can be
inferred that an exceptionally large pressure depletion accelerates
the collapse substantially. In this case, the bounce occurs at $t\sim
2.42\,t_{dyn}$, almost 3 times earlier. Despite that, the core still
bounces at approximately the same density and halts the
collapse. Following cooling, the remnant eventually settles into a
stable NS configuration. Similar to the previous set of simulations,
the end product is independent of whether the SI or SII cooling
strategy is adopted.

\subsubsection{Velocity perturbation}
Given that the binding energy of the TS is larger than that of the NS
with the same rest mass, it is possible that we must inject energy
into the system to overcome the bounce.  One way to induce a more
violent implosion is by introducing an inward velocity
perturbation. This perturbation is modeled with
Eq.~\eqref{velocity_pert}, which consists of two free parameters
$\xi_v$ and $\kappa$ to be specified.  We fix $\xi_v=-0.1$, which
corresponds to an inward velocity at the surface at 10\% of the speed
of light, and perform simulations with three different values of
$\kappa$, namely $\kappa = \{ 0.5,1,5 \}$. Fig.~\ref{kappa} shows the
result of these simulations in the evolution plot of the maximum
density, represented by maroon, green, and magenta colors,
respectively.

To explore a more diverse range of initial conditions, each of these
simulations is performed in two cases: $\xi_p=-0.01$ (dotted curves)
and $\xi_p=-0.90$ (solid curves).
In all these simulations, cooling is activated from the beginning,
that is, we employ the SII cooling strategy. All these cases yield the
same final result as the previous simulation of the collapsing WD; the
original bounce near the last stages of the collapse is still
unavoidable, and the remaining core is still a stable NS, regardless
of the value of $\kappa$. Different values of $\kappa$ can change only
the timescale of the collapse. Among these three choices, $\kappa=0.5$
and $\kappa=5$ result in the fastest and the slowest processes,
respectively. Moreover, the combination of the velocity perturbation
with a larger pressure depletion in the initial data leads to a
faster implosion in all cases.

\begin{figure}[t]
\includegraphics[width=\linewidth]{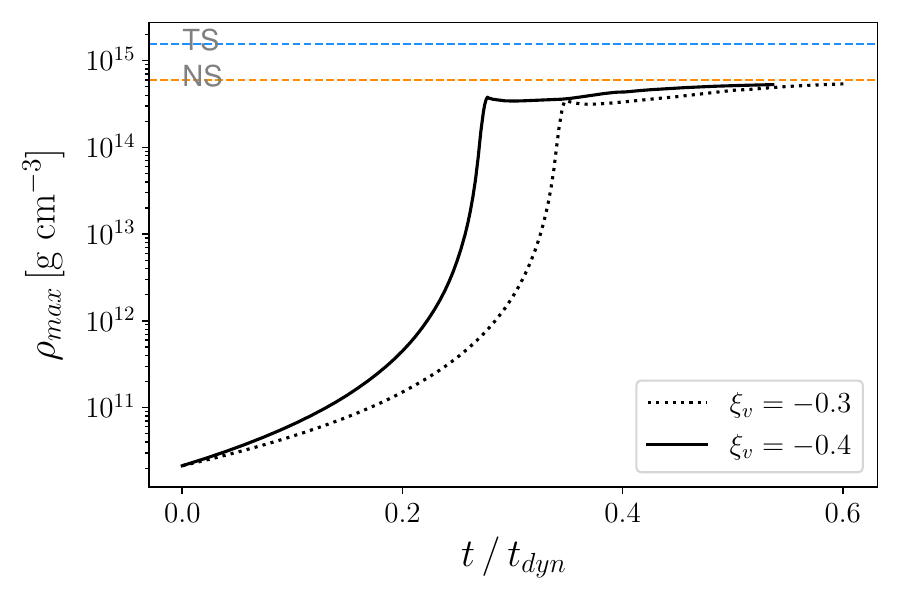}
\caption{\label{velocity_amplitude} The evolution of the maximum
  density of the collapsing WD with the most extreme initial
  conditions in pressure and velocity. The dotted curve shows the
  simulation with $\xi_v=-0.3$, and the solid curve represents the
  simulation with $\xi_v=-0.4$. Both the cases are performed in SII,
  with $\kappa=1$ and $\xi_p=-0.90$. Similar to Fig.~\ref{WDcollapse},
  the dashed lines show the maximum density of the stable NS/HS, and
  $t_{\rm dyn}$ is the dynamical timescale of the initial WD.}
\end{figure}

As the last set of simulations for the collapse of an unstable WD, we
also explore velocity perturbations with larger values of
$\xi_v$. Fig.~\ref{velocity_amplitude} presents these simulations with
$\xi_v=-0.3$ (dotted curve) and $\xi_v=-0.4$ (solid curve). Given that
all choices of $\kappa$ led to the same ultimate result, we run
these simulations only for the case of $\kappa=1$. Both simulations have 90
percent of pressure depletion, $\xi_p=-0.90$, and are conducted in the
SII\@. These represent the most extreme initial conditions examined in
this study. However, the result is once again the stable NS at
configuration \textbf{C}. The bounce still manifests after the
exponential collapse, although the process is even faster than in the
previous cases. The general outcome of these simulations suggests that
even extreme pressure and velocity perturbations cannot change the
eventual fate of the collapsing WD\@. The outcome is always a NS.

\subsection{Mass loss}
\label{mass_loss}
In this section, we investigate the scenario where the NS maximum mass
is somewhat lower than the Chandrasekhar limit so that a stellar core
or a WD at the Chandrasekhar limit collapses to the stable hybrid star
branch above the twin star mass regime. If this initial hybrid star
experiences some mass loss, e.g., due to strong winds, it could settle
into a TS instead of a NS.

\begin{figure}[h]
\includegraphics[width=\linewidth]{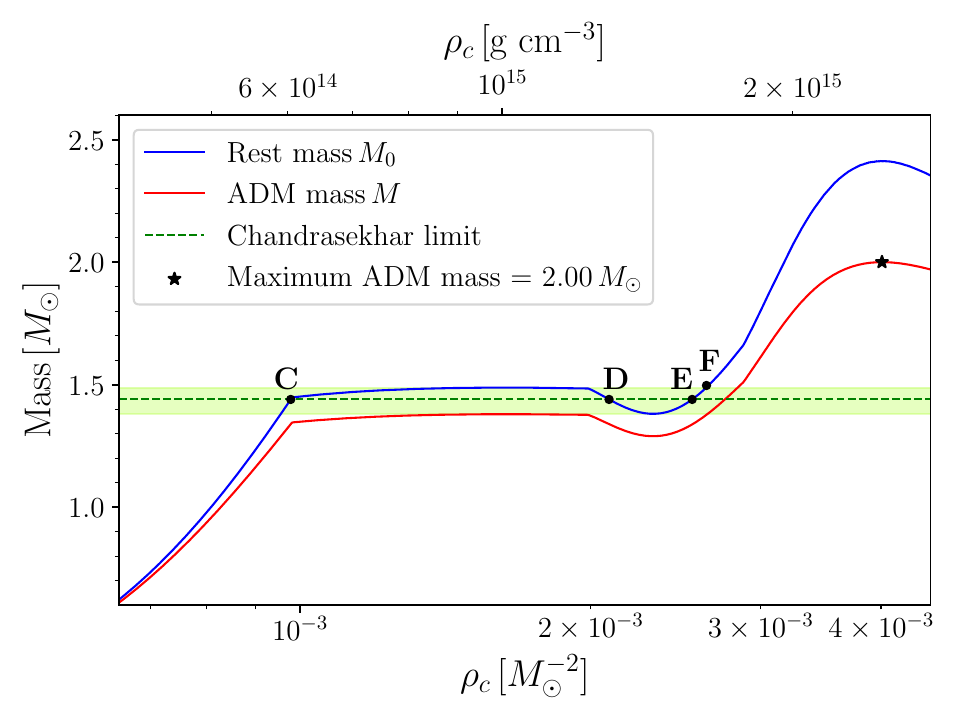}
\caption{\label{fig:shortEOS} An enlarged view of the NS/HS part of Fig.~\ref{fig:wideEOS}. Configuration \textbf{F} is a stable HS
  with a mass higher than the other three configurations
  indicated. The green shaded area roughly denotes the mass range for
  TSs. The red (blue) curve corresponds the ADM (rest) mass.}
\end{figure}

To model this scenario, we take the hybrid star configuration
\textbf{F} in Fig.~\ref{fig:shortEOS}, and deplete a small fraction of
its mass as described in Eq.~\eqref{mass_pert}. The initial rest mass
of this configuration is $M_0=1.49 \, M_\odot$, and exceeds
$M_{ch}$. The initial central rest-mass density is $\rho_c=1.63\times
10^{15} \, \textrm{g}\, \textrm{cm}^{-3}$ and the radius is $R=11.02
\, \textrm{km}$. By setting $\xi_m=-0.033$, the rest mass of the star
becomes $M_0 \approx M_{0,s}$, aligning with the rest masses of stable
stars \textbf{A}, \textbf{C}, and \textbf{E} in
Fig.~\ref{fig:wideEOS}. This initial perturbation is fairly strong
compared to a slow loss of mass through winds, but it allows us to get
a glimpse into whether this pathway to forming TSs is viable. We
checked the constraint violations after the mass loss remain small
throughout the simulations we performed (see discussion in
App.~\ref{sec:B}).

\begin{figure}[h]
\includegraphics[width=\linewidth]{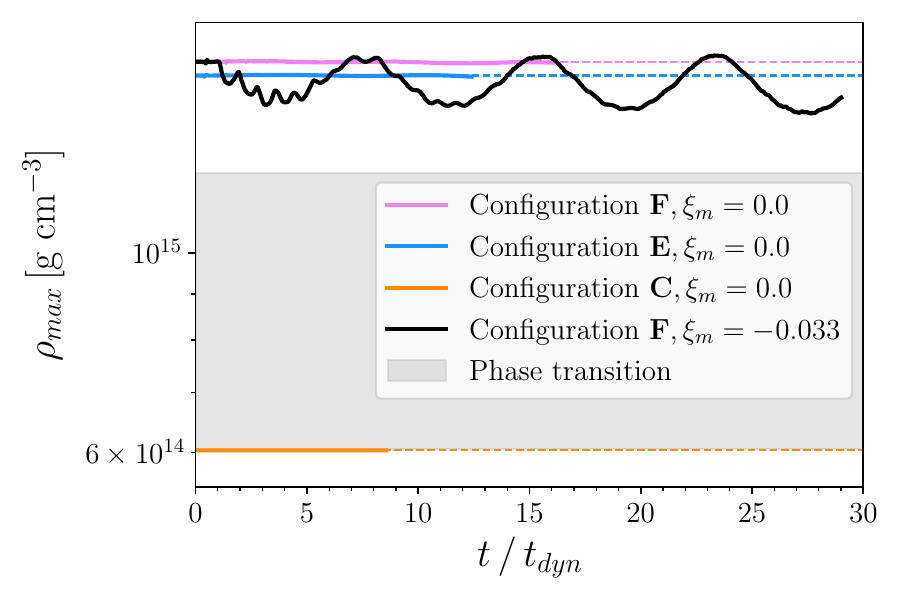}
\caption{\label{MassiveHS} Time evolution of the maximum density of
  configuration \textbf{F} without any perturbation (violet) and with
  a 3.3 percent mass loss (black). The stable NS and HS configurations
  \textbf{C} and \textbf{E} are shown in orange and blue,
  respectively. For each stable star, the solid line denotes the
  actual simulation, while the dashed line represents the initial
  central density. Furthermore, the shaded area indicates the density
  range of the phase transition. In this plot, $t_{\rm dyn}$ is the
  dynamical timescale of the initial HS.}
\end{figure}

\begin{figure*}
\includegraphics[width=\linewidth]{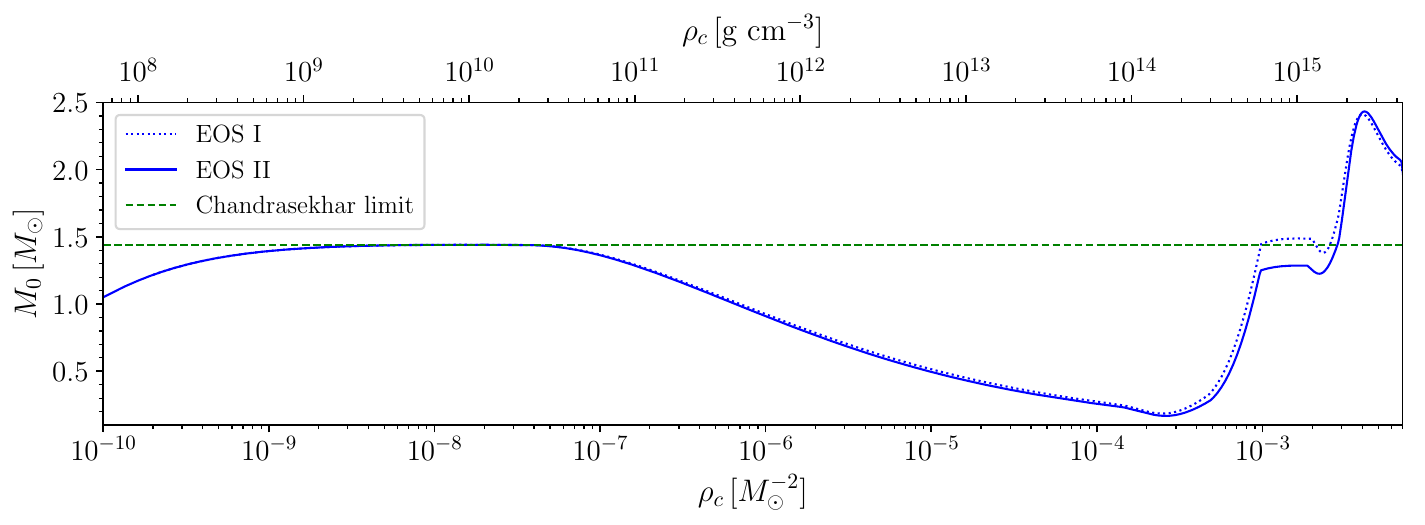}
\caption{\label{TOV_EOS_comparison} Rest mass as a function of central
  density for a sequence of TOV stars with EOS I (dotted line) and EOS II
  (solid line). The dashed line represents the Chandrasekhar mass. Notably,
  there is no TS in EOS II with a rest mass close to $M_{ch}$. In the
  density range relevant to WDs, the two curves exhibit almost
  complete overlap. Thus, the initial unstable WD configuration
  (\textbf{B}) shares similar characteristics in both EOSs.}
\end{figure*}

Fig.~\ref{MassiveHS} illustrates the results. Four simulations are
shown in this figure; the stable configurations \textbf{C},
\textbf{E}, and \textbf{F} (without perturbation) are displayed by the
orange, blue, and violet solid lines, respectively. The black solid
curve represents the evolution of the maximum density of a star
beginning at configuration \textbf{F} but perturbed by
$\xi_m=-0.033$. Interestingly, this plot demonstrates that the last
scenario successfully ends up with the formation of a stable HS in the
twin star regime. As a result of this large initial perturbation, two
types of oscillations can be seen around the maximum density of the
stable HS shown in blue. The first type exhibits oscillations with a
period of $\sim t_{dyn}$, which gradually disappears after a few
dynamical timescales. The second type of oscillation has a larger
domain with $\sim \pm 10 \, \%$ of deviations around the mean value of
the central rest-mass density of the TS \textbf{E}. The latter type of
oscillation has been seen before in simulations with EOSs that exhibit
a strong phase transition~\cite{Espino:2021adh,Hanauske:2018eej}, and
is likely due to a significant portion of the mass of the star
transitioning in and out of the density range in the hadron-to-quark
phase transition, where the stiffness of the EOS changes rapidly
leading to collapse (EOS softens) and bounce (EOS stiffens). We expect
that a gradual loss of rest mass would make such oscillations
disappear, but the main result of the final configuration being a TS
with the same rest mass should hold.  The important finding here is
that the maximum central density remains above the phase transition
region as shown in Fig.~\ref{MassiveHS}, and its mean is that of the
TS with the same rest mass.

\subsection{Hybrid star formation}

In the previous subsection, we demonstrated that TSs can form via mass
loss from a more massive HS. However, this process is plausible only
if a massive HS can form in the first place. In this subsection, we
explore the formation of a HS with a higher mass than TSs through 
gravitational collapse.

The simulation we perform here is the only one in this work based on
EOS II listed in Table~\ref{EOS_table}. Fig.~\ref{TOV_EOS_comparison}
compares the $M_0-\rho_c$ curves resulting from these two EOSs. The
TOV sequence resulting from EOS II has a NS peak lower than
$M_{ch}$. Therefore, there are no TSs with rest mass $M_{0,s}$.

\begin{figure}[t]
\includegraphics[width=\linewidth]{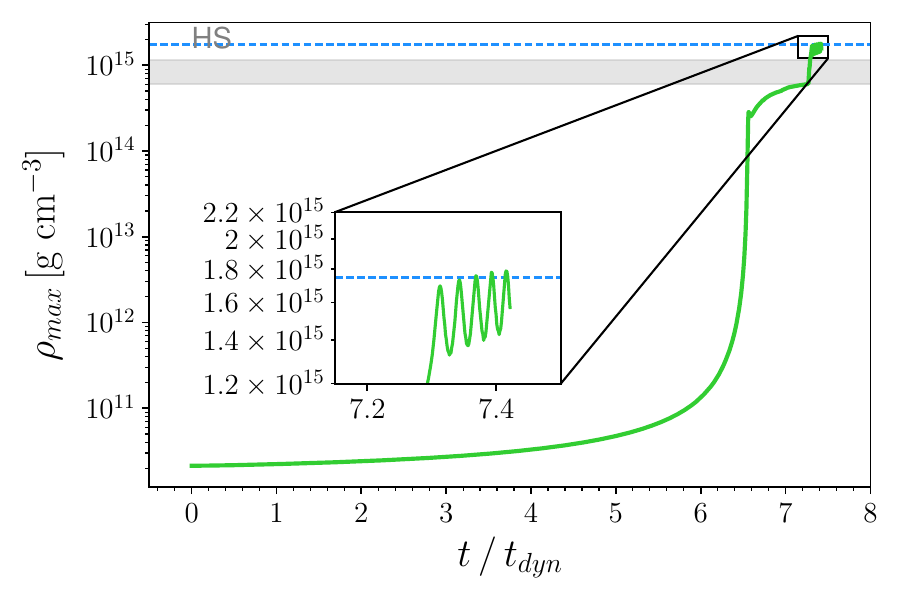}
\caption{\label{HS_formation} The evolution of the maximum density of
  a star beginning as an unstable WD with a rest mass
  $M_0=M_{0,s}$. The blue line represents the central density of the
  stable HS with this rest mass, and the phase transition range of
  densities is indicated by the shaded gray area. Here $t_{\rm dyn}$ represents
  dynamical timescale of the initial WD. The inset enlarges the final stages of collapse, indicating the convergence of maximum density to the central density of the stable HS.}
\end{figure}

Starting from an unstable WD with rest mass $M_{0,s}$ as initial data,
there should be only one stable equilibrium, cold configuration at
higher density. This equilibrium should correspond to a HS with a
central density of $\rho_c=1.75\times 10^{15} \, \textrm{g}\,
\textrm{cm}^{-3}$ and a radius $R=10.35 \, \textrm{km}$. The results
of this simulation are displayed in Fig.~\ref{HS_formation}, where we
show the evolution of the maximum rest-mass density. The blue line
denotes the described stable HS, and the shaded area shows the range
of densities in the phase transition. As expected, the remnant of
the collapse is this HS, as evidenced by the convergence of the
maximum density to the central density of this star.

The bounce observed in the previous simulations persists in this
simulation as well, due to the stiff EOS. SII cooling is employed in
this simulation, starting at $\tau_c=3\, t_{dyn}$. When the maximum
density approaches the phase transition segment of the EOS, the
evolution of the maximum density becomes very slow. Since we have
already confirmed that the cooling timescale does not affect the
outcomes, we change the cooling timescale to $\tau_c=1.5\, t_{dyn}$ at
$t/t_{dyn}\sim6.7$ to accelerate the transition. As the maximum
density enters the segment of the EOS that softens, the configuration
undergoes fast collapse to the HS formation. The only perturbation
introduced in this simulation is $\xi_p=-0.01$, applied solely to
speed up the initial collapse.

While not surprising, our simulations here demonstrate explicitly that
forming HSs with masses higher than TSs is viable.

\section{\label{sec:conclusion}Conclusions}

The exact form of the cold EOS beyond the nuclear saturation density
is still quite uncertain. Cores of NSs are a common place in nature
where neutron-rich matter can be found in this regime. The properties
of compact stars are intricately linked to the form of the EOS at
these densities. If a strong hadron-to-quark phase transition occurs
at these compact star densities, the EOS can potentially give rise to
a third branch of stable compact objects that consist of a quark core
surrounded by a hadronic shell. These configurations are known as
hybrid hadron-quark stars or hybrid stars. Associated with a third
family of compact stars is a range of masses where for each NS there
is a corresponding HS with the same mass but a smaller radius. These
stars are referred to as twin stars.

In this work, we explored the formation of TSs by conducting
hydrodynamical simulations in full general relativity. We constructed
piecewise polytropic EOSs that approximate realistic EOSs with a
hadron-to-quark phase transition, and we extended them down to the EOS
of a white dwarf.  Using our EOSs, we performed multiple simulations
of the collapse of an unstable WD under different initial conditions
with varying degrees of extremality of initial perturbations in
pressure and/or velocity. Given that total rest mass is conserved, the
ultimate product of the collapse could be either a stable NS or a
stable HS\@. The general finding from all our simulations is that the
unstable WD always collapses into a stable NS, regardless of how
extreme the initial conditions are. This overall result remains the
same even under extreme initial perturbations.

If TSs exist, there should be at least one pathway for them to
form. Following the standard theory of NS formation, a natural path
could be the formation of a more massive HS that subsequently loses a
small amount of mass in the form of winds. Another potential avenue
for mass loss would be a ``grazing'' collision of a massive HS with a
black hole where the massive HS undergoes an episode of mass loss and
then flies away. Such a scenario might take place in a dense stellar
cluster. Our simulations show that forming HSs heavier than in the
twin star mass regime is possible, and that forming a TS through a
heavier HS experiencing mass loss is a viable path. However, even this
scenario appears that it would require some fine-tuning for TSs to
form, which would place strong limits on the abundance of TSs. The
fine-tuning for the formation of twin stars results because of the
limited range of masses over which they are predicted to reside by
existing EOSs with hadron-to-quark phase transitions, and, hence, they
should be fairly rare objects. This conclusion and the challenge in
producing TSs as demonstrated by our simulations suggest that if a HS
star was involved in GW170817, then it likely was not a twin star.

We point out that our work is idealized in several ways and has a
number of caveats. First, we do not treat realistic neutrino effects
or magnetic fields, and the perturbations studied here represent a
simplified version of processes that may occur inside the dense core
of massive stars during the last stages of their evolution. In
particular, scenarios involving mass loss due to winds and mass gain
resulting from the infall of matter from outer shells of the initial
progenitor are expected to happen in nature. The results of our
simulations indicate that these perturbations play a notable role in
the potential formation of stable TSs, assuming that they exist in
nature. Thus, it is important to treat these
self-consistently. Nevertheless, our conclusion that NSs tend to be
the preferred outcome in the twin star mass range should hold, not
only because this is what our simulations demonstrate, but also
because of the very small range of masses over which twin stars are
predicted to exist by current EOSs, which, in turn, requires a delicate
balance of mass loss and mass gain.

To develop a more comprehensive model for the formation of TSs,
further cases should be studied with a wider range of parameters. Our
simulations were performed using static configurations. Rotation is
another aspect that can be added with different models to future
simulations to study the same problem. Additionally, these simulations
were conducted considering only one type of EOS\@. Neutrino effects,
nuclear reaction networks, and a more realistic progenitor are other
ways to increase the realism of these simulations.

The effects of rotation was not treated here, but such
  effects on HSs have been studied~\cite{Ayvazyan:2013cva,
    Largani:2021hjo, Rather:2021yxo, Tsaloukidis:2022rus}. The
  centrifugal barrier due to rotation is likely to lower the density
  compared to a nonrotating configuration for a given rest mass,
  thereby making matter stay well below the hadron-to-quark phase
  transition density. In addition, for some hybrid EOSs rotation has
  the effect of increasing the neutron star maximum mass higher than
  that of the third family~\cite{Bozzola:2019tit}. However, these
  results hold for cold configurations.
  
Temperature is another quantity that could affect
  stellar properties~\cite{Miao:2006eh}; notably, phase transitions
  may occur at finite rather than zero
  temperature~\cite{Carlomagno:2024vvr}.  The EOS we adopted in this
  work exhibits a hadron-to-quark phase transition at zero
  temperature. One scenario that can potentially change the results is
  the possibility of an EOS with a phase transition at finite
  temperatures~\cite{Baym:2017whm}. In this scenario, phase transition
  occurs at a lower density when the star is still
  hot~\cite{Bombaci:2016xuj}. Therefore, one can also think about the
  formation of TSs considering such a EOS with a finite-temperature
  phase transition. We will explore this possibility in future work.

Another potential pathway to forming HSs involves mergers of binary
NSs or WD-NS~\cite{Paschalidis:2009zz,Paschalidis:2011ez,Bauswein:2020ggy}\@. Exploring
all these scenarios and including more realistic physics is left for
future work. Most work in the area of TSs revolves around either
observational constraints on or properties of the EOS\@. Our work here
points out that there is another important problem surrounding TSs:
their formation and abundance in nature.




\begin{acknowledgments}
We thank Thomas Baumgarte for useful discussions.  We are grateful to
Zachariah Etienne and Leo Werneck for releasing their work on the
\texttt{IllinoisGRMHD} code which supports piecewise polytropic EOS,
and to all the \texttt{Einstein Toolkit} developers and
maintainers. We also thank Pedro Espino for sharing his EOS tables,
and D. Alvarez-Castillo, D. Blaschke and A.  Sedrakian for giving us
permission to use the ACS-II equations of state developed
in~\cite{Paschalidis:2017qmb}. Our Python TOV solver for tabulated
equations of state was developed starting from a code originally
created by Nikolaos Stergioulas. Simulations were performed on the
\texttt{Puma} cluster at the University of Arizona, and on ACCESS
resources under Allocation No. PHY190020. This work was supported in part
by NSF Grant No. PHY-2145421 and NASA Grant No. 80NSSC24K0771 to the
University of Arizona. G.\ B.\ was in part supported by the Frontera
Fellowship by the Texas Advanced Computing Center (TACC). Frontera is
funded by NSF Grant No.\ OAC-1818253.
\end{acknowledgments}

\appendix
\section{\label{sec:EOS_app}Extended EOS information}

The EOSs used in this work consist of eight segments and must satisfy the
list of conditions in Sec.~\ref{sec:EOS}. Empirically we have found
that each $\Gamma_i$ controls specific characteristics of the TOV
sequence shown in Fig.~\ref{fig:wideEOS}. $\Gamma_1$ primarily affects
the maximum WD mass, $\Gamma_2$ alters the density and mass of the WD
peak, $\Gamma_i$ for $i=\{ 3,4,5\}$ controls the maximum NS mass, and
$\Gamma_6$ changes the twin star mass range. Finally, $\Gamma_i$ for
$i=\{7,8\}$ affects the maximum HS mass. However, it is important to
emphasize that each of the free parameters affects more than just one
characteristic.

\begin{figure}[h]
\includegraphics[width=\linewidth]{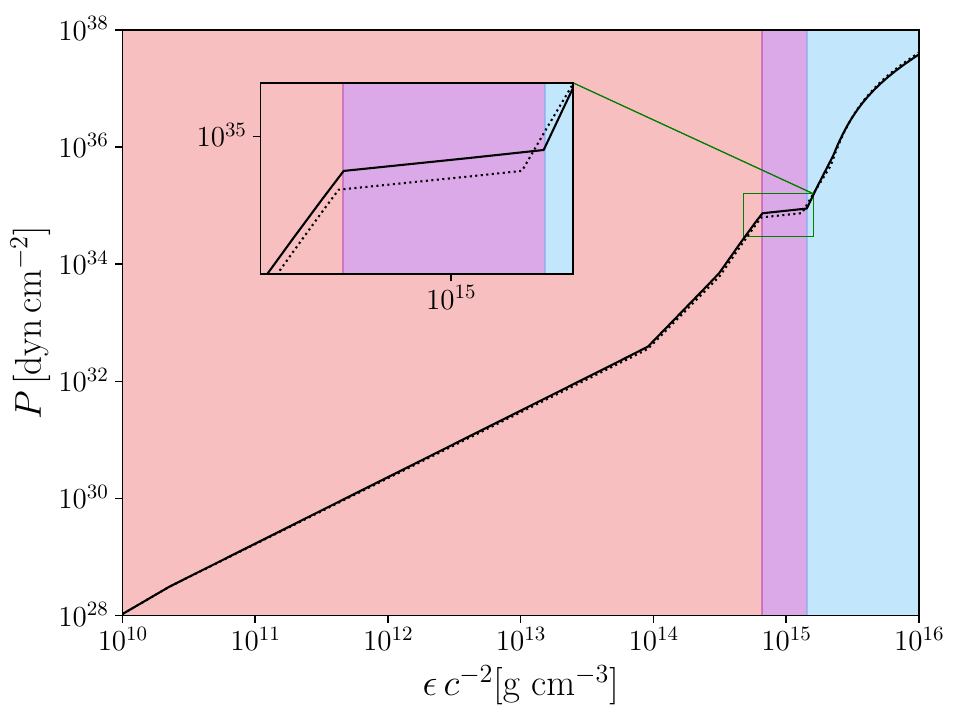}
\caption{\label{fig:eos}The relationship between the pressure and the
  energy density based on the EOSs constructed in
  Table~\ref{EOS_table}. EOS I is represented by the solid line, while
  the dotted line denotes EOS II. The shaded red, blue, and purple
  areas correspond to the pure hadronic, pure quark, and the mixed
  quark-hadron phases for EOS I, respectively. The inset enlarges
  the phase transition region.}
\end{figure}

The two EOSs listed in Table~\ref{EOS_table} exhibit only small
differences, which are mostly noticeable at higher
densities. Fig.~\ref{fig:eos} illustrates the pressure as a function
of energy density $\epsilon$ for our EOSs. The shaded red region
designates the density range where matter exists in the hadronic
phase, and the blue region shows the density range where it exists
purely in quark form. The narrower purple range, where the two regimes
overlap ($i=6$ branch of the EOS), is where the phase transition
occurs. In this regime, the original polytropic exponent was
$\Gamma_6=0$, corresponding to a first-order phase transition, but we
modified it to $\Gamma_6 = 0.2576$ to avoid a zero sound speed. This
implies that, instead of a sharp boundary from the hadron to the quark
phase, there exists a mixture of hadrons and quarks in this narrow
region of our EOS\@. This modification makes the sound speed of this
region $\simeq 15 \, \%$ of the speed of light, which allows for
stable numerical integration of the general relativistic hydrodynamic
equations.

\section{\label{sec:cooling} Cooling formalism}
In this appendix, we summarize the cooling formalism adopted in our
work.

In the standard 3+1 decomposition formalism to numerically solve the
Einstein equation, the spacetime metric $g_{\mu\nu}$ is written as follows
by~\cite{Arnowitt:1962hi}
\begin{equation} \label{metric}
ds^2 = -\alpha^2dt^2 + \gamma_{ij} (dx^i+\beta^i dt)(dx^j+\beta^j dt) \; ,
\end{equation}
where $\alpha$, $\beta^i$, and $\gamma_{ij}$ are the lapse function,
shift vector, and spatial metric, respectively ($i$ and $j$ are spatial
indices). For a perfect fluid with 4-velocity $u^\mu$, the
stress-energy tensor is written as
\begin{equation} \label{stress_tensor}
T^{\mu\nu}=\rho_0 h u^\mu u^\nu + Pg^{\mu\nu} \; .
\end{equation}
Here, the specific enthalpy $h$ is related to the specific internal energy $e=\frac{\epsilon}{\rho_0}-1$ through
\begin{equation} \label{enthalpy}
h=1+e+\frac{P}{\rho_0} \; .
\end{equation}
The evolution equations for general relativistic hydrodynamics (GRHD),
are given by~\cite{Duez:2005sf}
\begin{equation} 
 \begin{gathered}
  \partial_t \rho_\ast + \partial_j (\rho_\ast v^j) = 0 \; , \\[1em]
  \partial_t \tilde{\tau}+\partial_i (\alpha^2 \sqrt{\gamma} \; T^{0i} - \rho_\ast v^i) = s \; , \\[1em]
  \partial_t \tilde{S}_i+\partial_j(\alpha\sqrt{\gamma} \; T^j_{\;\;i}) = \frac{1}{2}\alpha\sqrt{\gamma} \; T^{\mu\nu} \; \partial_i g_{\mu\nu} \; . \\[1em]
  \label{GRMHD}
 \end{gathered}
\end{equation}
In these equations, $v^i=u^i/u^0$ is the coordinate three-velocity,
and the conservative variables are defined as
\begin{equation} 
 \begin{gathered}
  \rho_\ast = -\sqrt{\gamma}\;\rho_0 n_\mu u^\mu \; , \\[1em]
  \tilde{S}_i = -\sqrt{\gamma}\; T_{\mu\nu} n^\mu \gamma^\nu_{\;\;i} \; , \\[1em]
  \tilde{\tau} = \sqrt{\gamma}\; T_{\mu\nu} n^\mu n^\nu \rho_\ast \; ,
  \label{conservative_quant}
 \end{gathered}
\end{equation}
 where $n^\mu=\frac{1}{\alpha}(1,-\beta^i)$ is the timelike unit vector normal to $t=$ const slices.

In Eqs.~\eqref{GRMHD}, the first equation corresponds to the baryon
number conservation. The time component of energy-momentum
conservation $\nabla_\nu T_\mu^{\;\;\nu}$ is expressed in the second
equation, where $s=-\alpha\sqrt{\gamma}\;T^{\mu\nu} \nabla_\nu n_\mu$
is a source term\@. The spatial components of the
energy-momentum conservation are encoded in the third
equation. 

The set of Eqs.~\eqref{GRMHD} also requires an EOS to close. The
general form of the EOS that is assumed to solve this set of equations
is given by Eq.~\eqref{cold_th}~\cite{Stephens:2008hu}, where
$P_{cold}$ stands for the pressure of the zero-temperature matter, and
has the polytropic form introduced in Sec.~\ref{sec:EOS}\@. Any
additional pressure due to heating, e.g., by shock heating, is
encapsulated in $P_{th}$. This term is considered as a $\Gamma$-law:
\begin{equation} \label{P_th}
P_{th}=(\Gamma_{th}-1)\rho_0 e_{th} \; ,
\end{equation}
where $e_{th}=e - e_{cold}$, and $e_{cold}$ is the specific internal energy associated with the cold pressure, which can be found from the first law of thermodynamics as
\begin{equation} \label{eps_th}
e_{cold}(\rho_0)=-\int P_{cold}(\rho_0)d\left(\frac{1}{\rho_0}\right) \; .
\end{equation}
For the specific case of Eq.~\eqref{piecewise}, this becomes
\begin{equation} \label{eps_th_poly}
e_{cold}=\frac{k_i}{\Gamma_i - 1}\rho_0^{\Gamma_i -1} + a_i \; ,
\end{equation}
where $a_i$ is a constant of integration. Considering $\epsilon_{cold}$ as the energy density when there is no heat, this constant is determined as~\cite{Read:2008iy}
\begin{equation} \label{integ_const}
a_i=\frac{\epsilon_{cold}(\rho_{0,i})}{\rho_{0,i}}-1-\frac{k_i}{\Gamma_i -1}\rho_{0,i}^{\Gamma_{i-1}} \; .
\end{equation}
In the presence of radiation, the conservation of energy-momentum must
be modified to incorporate the radiation stress-energy tensor
$R^{\mu\nu}$:
\begin{equation} \label{modified_EM}
\nabla_\mu (T^{\mu\nu}+R^{\mu\nu})=0 \; .
\end{equation}
The dynamics of the radiation stress-energy tensor is described by
$\nabla_\mu R^{\mu\nu}=-G^\nu$, where $G^\mu$ is the radiation four-force
density.  Assuming local and isotropic (in a frame comoving with the fluid) cooling, the resulting energy and momentum equations are then given by~\cite{Paschalidis:2011ez}
\begin{equation} 
 \begin{gathered}
  \partial_t \tilde{\tau}+\partial_i (\alpha^2 \sqrt{\gamma} \; T^{0i} - \rho_\ast v^i) = s -\alpha^2 \sqrt{\gamma}\; u^0 \Lambda \; , \\[1em]
  \partial_t \tilde{S}_i+\partial_j(\alpha\sqrt{\gamma} \; T^j_{\;\;i}) = \frac{1}{2}\alpha\sqrt{\gamma} \; T^{\mu\nu} \; \partial_i g_{\mu\nu} - \alpha \sqrt{\gamma}\; u_i \Lambda \; ,
  \label{cooling_GRMHD}
 \end{gathered}
\end{equation}
where we choose the same emissivity $\Lambda$ as in~\cite{Paschalidis:2011ez} 
\begin{equation} \label{Lambda}
\Lambda = \frac{\rho_0}{\tau_c}e_{th} \; .
\end{equation}
Here $\tau_c$ is the cooling timescale. 

Cooling changes the internal thermal energy over time. The governing equation in a comoving frame can be written as~\cite{Paschalidis:2011ez}
\begin{equation} \label{cooling_eps}
\frac{de_{th}}{d\tau}=\left[\frac{(\Gamma_{th}-1)}{\rho_0}\frac{d\rho_0}{d\tau}-\frac{1}{\tau_c}\right]e_{th} \; ,
\end{equation}
where $\tau$ is the proper time. When no adiabatic contraction or expansion takes place, then $\rho_0$ is constant, and the last equation describes an exponential evolution of $e_{th}$ as a function of proper time, that is
\begin{equation} \label{exponential}
e_{th} \propto \exp{\left(-\frac{\tau}{\tau_c}\right)} \; .
\end{equation}

\section{\label{sec:A}Cooling tests}

To validate our implementation of cooling we introduce a uniform
perturbation in pressure everywhere in a test simulation involving a
stable NS\@. We adopt a cold polytropic function EOS, $P=k
\rho_0^\Gamma$, where $\Gamma=2$ and
$k=100\,M_\odot^2$. Equation~\eqref{cold_th} implies that any positive
perturbation in pressure adds some $P_{th}$ to the initial cold
pressure. We conduct three different simulations for a stable NS with
the central density $\rho_c=0.001\,M_\odot^{-2}$. In the first
simulation, no pressure perturbation is applied, and cooling remains
inactive. In the second simulation, we impose a small pressure
perturbation with $\xi_p=0.01$, which consequently, produces heat, but
cooling is not active. Finally, we conduct a simulation that includes
the same pressure perturbation as the second simulation, while cooling
is active with $\tau_c=3.16\,t_{dyn}$. We continue the simulations for
$\sim 30\, t_{dyn}$.

\begin{figure*}
\includegraphics[width=\linewidth]{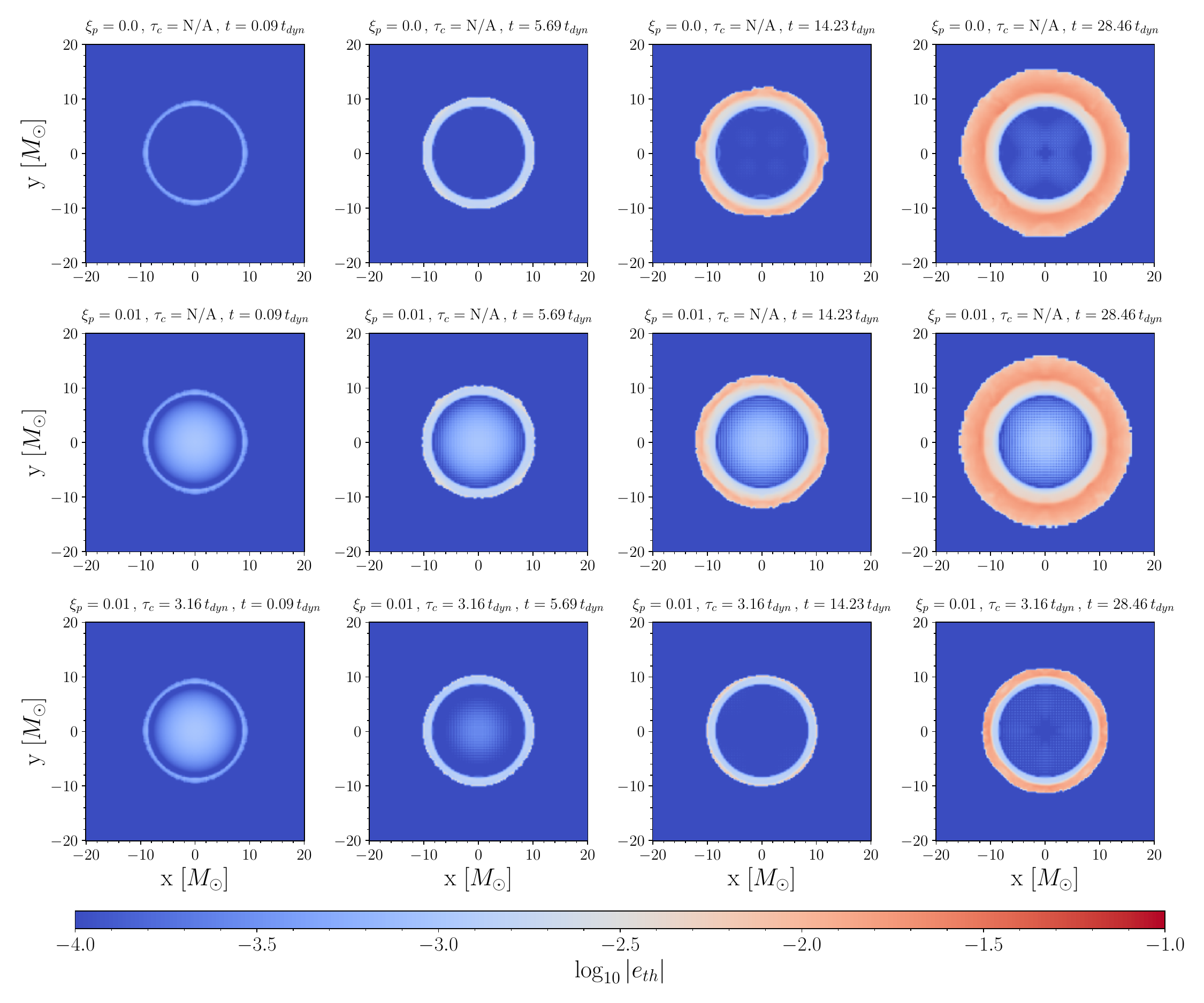}
\caption{\label{cooling} The value of $|e_{th}|$ is shown in
  two-dimensional snapshots of three simulations involving a stable NS
  governed by a $\Gamma=2$ polytropic EOS with constant
  $k=100\,M_\odot^2$, and $\rho_c=0.001\,M_\odot^{-2}$ . The evolution
  of each simulation is illustrated in a separate row. The first row
  describes a NS without any initial perturbation in pressure and
  without cooling. The second row corresponds to the case with a small
  pressure perturbation in the initial data and no cooling. The bottom
  row has the same initial data as the second row but cooling is
  active. The plot demonstrates that cooling is effective at removing
  excess heat in the bulk of the stellar matter. The stellar surface
  discontinuity generates heat at every time step and would require
  more aggressive cooling to completely cool. However, the amount of
  mass in that hotter region is negligible compared to that of the
  cold bulk of the star. }
\end{figure*}

Fig.~\ref{cooling} shows the result of these simulations for the
absolute value of $e_{th}$ at four different times. The three
simulations are presented in the three rows, respectively. Even in the
first simulation, where there is no additional thermal pressure in the
initial data, small numerical errors result in the generation of some
heat that grows over time. These numerical errors can also deplete
heat, leading to negative $e_{th}$ at some points. This is why we show
the absolute value of this quantity. The initial data for this
simulation contain no initial heat, as there is no heat source of
$P_{th}$ at the beginning. In the second simulation, the perturbation
$\xi_p=0.01$ results in heat in the initial data. This excess heat can
accumulate alongside the contribution from the numerical errors over
time, resulting in a relatively hotter configuration. The third
simulation starts with the same amount of initial heat as the second
simulation, but cooling removes the extra heat from the star. As can
be seen in the bottom row of plots in Fig.~\ref{cooling}, the initial
heat and a significant portion of the heat arising from numerical
errors are effectively removed after a few dynamical times.

\begin{figure}[t]
\includegraphics[width=\linewidth]{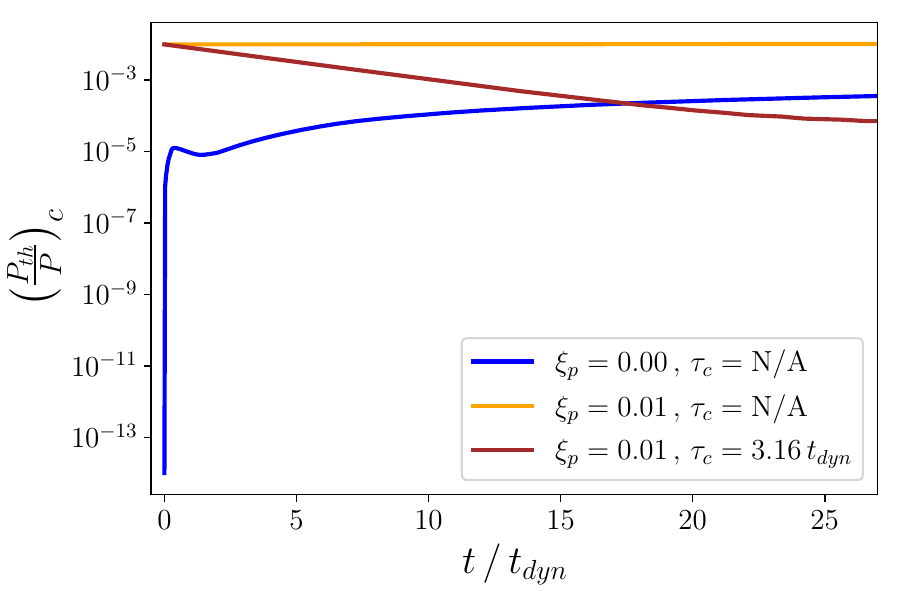}
\caption{\label{cooling_pressure_test} The time evolution of
  $P_{th}/P$ at the center of the star is shown for three different
  simulations. The blue curve corresponds to a simulation without any
  perturbation or cooling. The orange curve has no cooling, but
  incorporates a pressure perturbation with $\xi_p=0.01$. The dark red
  curve corresponds to the simulation with $\xi_p=0.01$ but cooling is
  active with $\tau_c=3.16\,t_{dyn}$.}
\end{figure}

\begin{figure}[h]
\includegraphics[width=\linewidth]{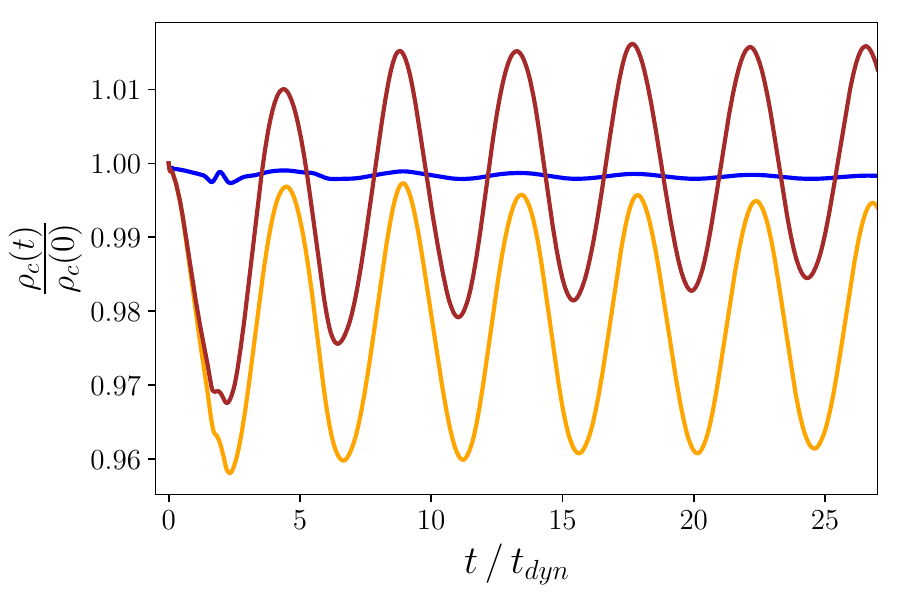}
\caption{\label{cooling_density_test} Time evolution of the central density for the three simulations shown in Fig.~\ref{cooling_pressure_test}, with colors having the same meaning.}
\end{figure}

We can also examine the evolution of $P_{th}$ at the center of the
star in these three simulation, as depicted in
Fig.~\ref{cooling_pressure_test}. In the first simulation, $P_{th}$
increases over time as a result of numerical error. The value of
$P_{th}$ remains constant over time in the second simulation, as
expected. In the last simulation, the cooling mechanism gradually
decreases the value of $P_{th}$. After $\sim 18\, t_{dyn}$, the
central value of $P_{th}$ in this simulation drops below its
corresponding value in both the other simulations, revealing the
successful removal of heat by cooling. The positive $\delta P$ also
leads to a decrease in density at every point. This behavior can be
seen in Fig.~\ref{cooling_density_test}. The positive perturbation
induces more pronounced oscillations in quantities such as density,
similar to the oscillations that will be discussed in
App.~\ref{sec:stable_stars} for stable stars. These oscillations
remain strong even after many $t_{dyn}$, while cooling has eliminated
the majority of the heat. In this, cooling drives the maximum
density to converge to its initial value and oscillate around it.

The results of these simulations indicate that cooling is successfully
removing excess heat, restoring the EOS to its cold state\@. Having
verified the functionality of the code, we test its accuracy. In other
words, cooling should operate as modeled in App.~\ref{sec:cooling}. To
design an actual (quantitative) cooling test, we may focus on the
evolution of the quantity ${e_{th}}\slash{e}$ at the center over the
very early times, for a simulation with $\xi_p=0.01$ and
$\tau_c=3.16\, t_{dyn}$. If $\rho_0$ is fixed, the value of $e_{th}$
must be decreasing as an exponential function with the proper time, as
expressed by Eq.~\eqref{exponential}. Unlike the previous plot, the
evolving quantity here is shown as a function of proper time $\tau$
rather than coordinate time $t$. This adjustment is made due to the
exponential behavior being dependent on proper time. To convert the
coordinate time to proper time, the value of the lapse function is
obtained at each timestep, and the conversion is achieved by

\begin{equation} \label{lapse}
d\tau=\alpha \, dt \; .
\end{equation}

Then, the result can be fitted to an exponential function characterized by a damping timescale $\tau_{damp}$ as

\begin{equation} \label{damp}
e_{th} \propto \exp{(-\frac{\tau}{\tau_{damp}})} \; .
\end{equation}

\begin{figure}[t]
\includegraphics[width=\linewidth]{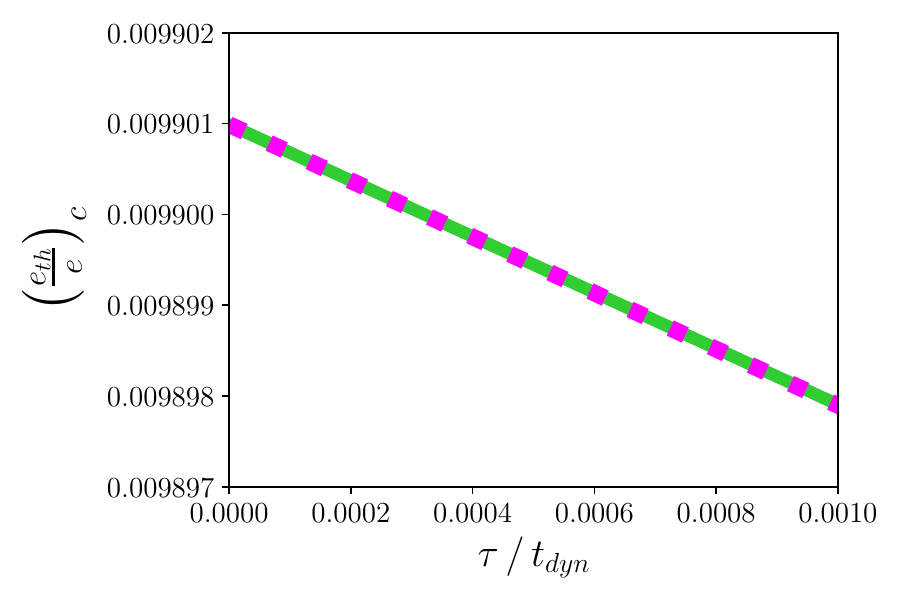}
\caption{\label{CoolingTest} The evolution of the quantity $e_{th}/e$
  at the center of a NS during the early times. The green curve
  represents the simulation results with $\xi_p=0.01$ and
  $\tau_c=3.16\, t_{dyn}$. The fitted model described by
  Eq.~\eqref{damp} is shown by dots, with the damping timescale set to
  $\tau_{damp}=3.20\, t_{dyn}$.}
\end{figure}

The solid curve in Fig.~\ref{CoolingTest} shows the simulation results,
and the dots represent the fitted model based on Eq.~\eqref{damp} with
$\tau_{damp}=3.20\, t_{dyn}$ to this curve at early times. Note that
$\tau_c=\tau_{damp}$ only if $d\rho_0/d\tau=0$. Although the density
changes only little in our simulations, it is not constant. Therefore
we expect that the measured $\tau_{damp}$ will deviate slightly from
the expected value of $\tau_c$. In our simulation, the
difference between $\tau_c$ and $\tau_{damp}$ is only $\sim 1 \%$.

\section{\label{sec:stable_stars}Stable stars}

In this appendix, we simulate the two stable configurations \textbf{C}
and \textbf{E} in Fig.~\ref{fig:wideEOS}, representing a NS and a HS,
respectively\@. We do this to demonstrate that our code can evolve
stable stars, and since we expect these two configurations to be the
end points for a collapsing unstable WD, i.e. configuration
\textbf{B}. Configuration \textbf{C} is the stable NS with a central
density of $\rho_c=6.04\times 10^{14} \, \textrm{g} \,
\textrm{cm}^{-3}$ and a radius $R=13.78 \, \textrm{km}$. Configuration
\textbf{E}, on the other hand, is a stable HS with a central density
of $\rho_c=1.57\times 10^{15} \, \textrm{g} \, \textrm{cm}^{-3}$ and a
radius $R=11.40 \, \textrm{km}$.

\begin{figure}[h]
\includegraphics[width=\linewidth]{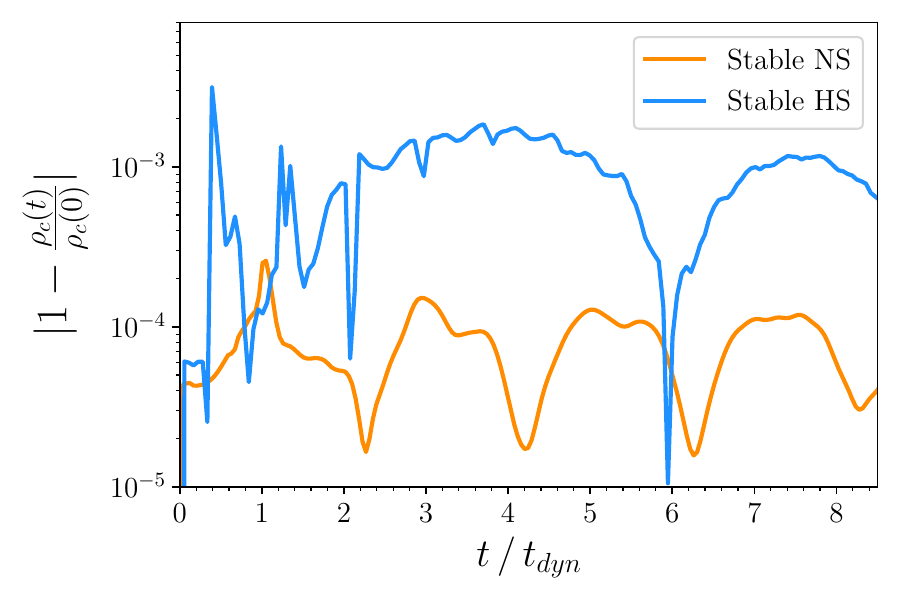}
\caption{\label{densityNSHS} Evolution of the central rest-mass
  density for the stable NS configuration \textbf{C} (orange) and for
  the stable HS configuration \textbf{E} (blue). The coordinate time
  $t$ is scaled with the dynamical timescale $t_{dyn}$, which depends
  on the initial maximum density $\rho_c(0)$ and is different for the
  two stars. The plot demonstrates that we can accurately evolve stable
  compact stars with the EOSs we constructed.}
\end{figure}

For these simulations, the grid structure consists of eight and seven
refinement levels, respectively. In both configurations, the finest
resolution is set to 100 points across the initial radius of the
stars. The half-side length of the coarsest box is $134.4 R$ for the
stable NS simulation and $67.2 R$ for the stable HS simulation.

Fig.~\ref{densityNSHS} shows the relative deviation of the central
density over time, with respect to its initial value. This figure
indicates that over $8\,t_{dyn}$, the stable NS exhibits density
fluctuations of $\sim 0.01 \%$ at maximum, while the largest amplitude
of fluctuations for the HS is $\sim 0.1 \%$. Because of the transition to
the more complex part of the EOS, the density evolution of the stable
HS follows a less predictable pattern.

The primary conclusion drawn from these simulations is that the stable
configurations of the TSs are dynamically stable indeed, and our code
can accurately evolve them.

\section{\label{sec:B}Constraints}
To ensure the accuracy of the evolution of the Einstein equations we
monitor the Hamiltonian and momentum constraint equations
(see~\cite{Baumgarte:2010ndz} for a detailed discussion on the 3+1
decomposition of Einstein's equations). This is important because after
perturbing our initial data, we do not resolve the constraints, so any
constraint violations should remain small. In this appendix, we
demonstrate that this is the case.

\begin{figure}[h]
\includegraphics[width=\linewidth]{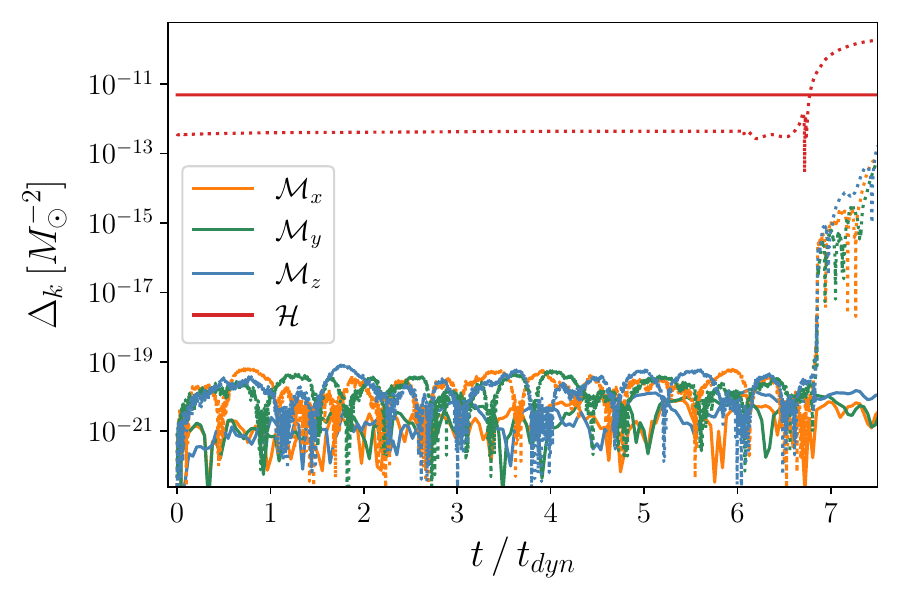}
\caption{\label{constraint} The constraints as a function of time
  scaled by the initial dynamical timescale for the Hamiltonian
  constraint (represented in red), and x, y, and z components of the
  momentum constraint (shown in orange, green, and blue,
  respectively). Solid curves stand for the simulation of the stable
  NS configuration \textbf{C}, while dotted curves denote the
  collapsing WD configuration \textbf{B}. The sudden rise in
  $\Delta_k$ when the initial WD enters rapid collapse is
  unavoidable. However, the value of $\Delta_k$ remains sufficiently
  small throughout the simulation.}
\end{figure}

The absolute value of the constraint violation averaged over volume as
a function of time is shown in Fig.~\ref{constraint} for both a stable
NS (solid curves) and a collapsing WD with cooling activated from the
beginning (dotted curves). The constraint violations are denoted by
$\Delta_k$.  Red corresponds to the violation of the Hamiltonian
constraint $\mathcal{H}$, while the momentum constraint violations,
$\mathcal{M}_j$ with $j=\{x,y,z\}$, correspond to orange, green, and
blue. $\Delta_k$ should remain close to zero at all time steps. As the
star undergoes collapse, the computational errors increase, leading to
larger values of $\Delta_k$. Since $\Delta_k$ is a dimensionful
quantity, we also show $\Delta_k$ for the stable NS simulation
discussed in App.~\ref{sec:stable_stars} for comparison. The plot
demonstrates that the constraints remain well satisfied throughout the
collapse. The maximum value of $\Delta_k$ remains small even in the
simulations with more aggressive initial perturbations presented in
this work, and it similarly occurs during the late stages of the
collapse. The worst case in our simulations involved the HS with mass
loss, where constraint violations are still $\Delta_k \lesssim
10^{-9}\,M_\odot^{-2}$.

\bibliography{apssamp}
\end{document}